\documentclass[12pt]{article}
\usepackage{amssymb,amsmath,bm}
\usepackage{threeparttable}
\usepackage{color}
\usepackage{enumitem}
\usepackage[colorlinks,linkcolor=blue]{hyperref}
\usepackage{geometry}
\usepackage{tabularx}
\usepackage{bm}
\usepackage{authblk}
\usepackage{nicematrix}
\usepackage{tikz}
\usepackage[square, comma, sort&compress, numbers]{natbib}
\usepackage{nicematrix}
\usepackage{float} 
\usepackage{multirow} 
\usepackage{booktabs} 
\usepackage{makecell}
\newcommand{\rank}{{\mathrm{rank}}}
\newtheorem{theorem}{Theorem}[section]
\newtheorem{definition}{Definition}[section]
\newtheorem{lemma}{Lemma}[section]
\newtheorem{example}{Example}[section]

\newtheorem{corollary}{Corollary}[section]
\newtheorem{remark}{Remark}[section]

\catcode`@=11 \@addtoreset{equation}{section} \catcode`@=12
\begin{document}
	\title{\bf A Class of Self-Orthogonal, NMDS, or LCD MDS $(\mathcal{L}, \mathcal{P})$-Twisted Generalized Reed-Solomon Codes  \thanks{ Email: liangzhongh0807@163.com; 3120193984@qq.com; huangdongmeimaths@stu.sicnu.edu.cn; qunyingliao@sicnu.edu.cn; tangchunmingmath@163.com.  This paper is supported by National Natural Science Foundation of China (Grant No. 12471494, Grant  No. 12231015) and Natural Science Foundation of Sichuan Province (2024NSFSC2051).}}
	\author[1]{\small Zhonghao Liang}\author[1]{\small { Chenlu Jia}}\author[2]{\small Dongmei Huang}
	\author[1]{\small Qunying Liao\thanks{Corresponding author: Qunying Liao.}
	}\author[3]{\small Chunming Tang}
	\affil[1] {\small College of Mathematical Sciences, Sichuan Normal University, Chengdu, 610066, China}
	\affil[2] {\small School of Mathematical Sciences, China West Normal University, Nanchong 637009, China.}
	\affil[3] {\small School of Information Science and Technology, Southwest Jiaotong University, Chengdu,
		611756, China}	
	\date{}
	\maketitle
	{\bf Abstract.}
	{\small 
Maximum distance separable (in short, MDS), near MDS (in short, NMDS), linear complementary dual codes (in short,  LCD) and self-orthogonal codes play a pivotal role in algebraic coding theory, particularly in applications such as quantum communications and secret sharing schemes.
Recently, the construction of non-generalized Reed-Solomon (in short, non-GRS) codes has emerged as a significant research frontier. This paper focuses on a class of $(\mathcal{L}, \mathcal{P})$-twisted generalized Reed-Solomon (TGRS) codes $\mathcal{C}_{k-1}\left(\boldsymbol{\alpha},\boldsymbol{v},\boldsymbol{\eta}\right)$ characterized by $\ell+1$ twists. By considering a special class of parity-check matrices derived from existing results, we derive the corresponding necessary and sufficient conditions or sufficient conditions for $\mathcal{C}_{k-1}\left(\boldsymbol{\alpha},\boldsymbol{v},\boldsymbol{\eta}\right)$ to be self-orthogonal, and then generalize the corresponding results given by Liu et al. (2021), Sui et al. (2021), and Zhang et al. (2025). As applications, leveraging these self-orthogonal structures, we can obtain several families of LCD MDS codes. Secondly, we provide a characterization of the NMDS property for the codes $\mathcal{C}_{k-1}\left(\boldsymbol{\alpha},\boldsymbol{v},\boldsymbol{\eta}\right)$, offering a partial solution to the open problem concerning general $(\mathcal{L}, \mathcal{P})$-TGRS codes posed by Hu et al. (2025). Finally, we rigorously prove that the codes $\mathcal{C}_{k-1}\left(\boldsymbol{\alpha},\boldsymbol{v},\boldsymbol{\eta}\right)$ are of non-GRS type when $2k > n>k+\ell+2$.


\section{Introduction}
 Let $\mathcal{C}$ be an $[n, k, d]$ linear code over the finite field $\mathbb{F}_{q}$ with $q$ elements and  $\mathcal{C}^{\perp}$ be the Euclidean dual code of $\mathcal{C}$. If $\mathcal{C}\cap\mathcal{C}^{\perp}=\left\{\boldsymbol{0}\right\}$, then $\mathcal{C}$ is linear complementary dual codes (in short,  LCD). If $\mathcal{C}\subseteq\mathcal{C}^{\perp}$, then $\mathcal{C}$ is self-orthogonal. Especially, if $\mathcal{C}=\mathcal{C}^{\perp}$, then $\mathcal{C}$ is self-dual. In addition, if $d=n-k+1$,
then $\mathcal{C}$ is maximum distance separable (in short, MDS). If
$d=n-k$, then  $\mathcal{C}$ is almost MDS (in short, AMDS). In particular, if both $\mathcal{C}$ and $\mathcal{C}^{\perp}$ are AMDS, then $\mathcal{C}$ is near MDS (in short, NMDS). If $\mathcal{C}$ is not equivalent to any generalized Reed-Solomon (in short, GRS) code, then $\mathcal{C}$ is called to be non-GRS type. 

In recent years, linear codes with some special properties have received renewed attention due to their important role
in new applications  \cite{BringerJ2014,Calderbank1997,CarletEHLCDMDS2018,Gottesman1996,Calderbank1998,ChenAGTIT2005,JinEHSOAGTIT2011}. For example,  self-orthogonal codes can be used to
construct  LCD codes  \cite{CarletEHLCDMDS2018} or quantum error-correcting codes, which can protect
quantum information in quantum computations and quantum
communications \cite{Calderbank1997,Gottesman1996}. Euclidean self-dual codes and NMDS codes can be used to find
diverse applications in cryptographic protocols (e.g. secret sharing schemes) and combinatorics \cite{Cramer2008,Dougherty2008,Simos2012,Zhou2009}. While, for many known linear codes, they are not necessarily self-orthogonal, self-dual, or NMDS, and so the corresponding characterization and construction are a very interesting problem \cite{HanDCDCC2024,WangSOAGarxiv2025,WangSOarxiv2025,LiuTECarxiv2025}.  

In 2017, Beelen et al. \cite{BeelenTGRS2017} first introduced twisted generalized Reed-Solomon (in short, TGRS) codes to construct non-GRS MDS codes, which are a generalization of classical GRS codes. Unlike GRS codes, TGRS codes are not necessarily MDS. Many scholars have investigated the MDS properties of TGRS codes, as well as their NMDS properties \cite{HuangTGRSDCC2021}, self-dual properties \cite{Zhu*TGRSCC2024,SuiTGRS2023CC,ZhangTGRSDCC2022}, self-orthogonal properties \cite{LuoHSOTGRSAMC2022,ZhuSoTGRSarxiv20222}, and other coding properties \cite{ChengTGRSHTIT2023,LiuTGRSDCC2021,WuTRS1HullTIT2020,HeECPTGRSDM2023,FangTRSFFA2025,JiaTGRSDCC2025,LiTWTGRSTIT2025}. In 2025, Zhao et al. \cite{ZhaoATGRSTIT2025} proposed arbitrary twisted generalized Reed-Solomon (in short, A-TGRS) codes, which extend the definition of TGRS codes, and further constructed several families of Hermitian self-dual A-TGRS codes \cite{ZhaoTGRSarxiv2025}. Recently, Hu et al. \cite{HuTGRSTIT2025} presented a more flexible definition of TGRS codes than that given in \cite{ZhaoATGRSTIT2025}, namely
	\[
	(\mathcal{L},\mathcal{P})\text{-}\mathrm{TGRS}_{k}(\mathcal{L},\mathcal{P},\boldsymbol{B})\triangleq\left\{\left(v_{1}f\left(\alpha_{1}\right), \ldots,v_{n}f\left(\alpha_{n}\right)\right) \,\bigg|\, f(x) \in \mathcal{F}_{n,k}(\mathcal{L}, \mathcal{P}, \boldsymbol{B})\right\},
	\]
	where $\mathcal{L}\in\left\{0,1,\ldots,n-k-1\right\}, \mathcal{P}\in\left\{0,1,\ldots,k-1\right\},\boldsymbol{B}=(b_{i,j})\in\mathbb{F}_{q}^{k\times (n-k)}$ with $0\leq i\leq k-1, 0\leq j\leq n-k-1$, and $\boldsymbol{v}=\left(v_{1},\ldots,v_{n}\right)\in\left(\mathbb{F}_{q}^{*}\right)^{n}$. Here,
	\[
	\mathcal{F}_{n,k}(\mathcal{L}, \mathcal{P}, \boldsymbol{B})=\left\{\sum_{i=0}^{k-1} f_{i} x^{i}+\sum_{i \in \mathcal{P}} f_{i} \sum_{j \in \mathcal{L}} b_{i, j} x^{k+j}: f_{i} \in \mathbb{F}_{q}, 0 \leq i \leq k-1\right\}.
	\]
	The code $(\mathcal{L},\mathcal{P})$-$\mathrm{TGRS}_{k}(\mathcal{L},\mathcal{P},\boldsymbol{B})$ is referred to as the $(\mathcal{L},\mathcal{P})$-TGRS code, and the matrix $\boldsymbol{B}$ is defined as its coefficient matrix.

 In the past few years, for some special $\boldsymbol{B}$, there have been many results  \cite{HuangTGRSDCC2021,Zhu*TGRSCC2024,SuiTGRS2023CC,ZhangTGRSDCC2022,LuoHSOTGRSAMC2022,ZhuSoTGRSarxiv20222,ChengTGRSHTIT2023,LiuTGRSDCC2021,WuTRS1HullTIT2020,HeECPTGRSDM2023,FangTRSFFA2025,JiaTGRSDCC2025,LiTWTGRSTIT2025,ZhaoATGRSTIT2025,ZhaoTGRSarxiv2025,HuTGRSTIT2025,ZhuTGRSFFA2024,YangTGRSFFA2025,MeenaKCTGRSDCC2025,ZhangTGRSJAA2025,DingTGRSAMC2025,SuiTGRSTIT2022,GuTGRSTIT2024,SuiTGRSDCC2022,LiangLCD*TGRS2026}. Especially, we list some results as follows:
\begin{itemize}
		\item In 2021, Huang et al. \cite{HuangTGRSDCC2021} completely determined the existence of self-dual codes for the $(\mathcal{L},\mathcal{P})$-TGRS code with coefficient matrix
		\[
		\boldsymbol{B}=
		\begin{pmatrix}
			\boldsymbol{0}_{(k-1)\times 1}&\boldsymbol{0}_{(k-1)\times n-k-1}\\ 
			b_{k-1,0} &\boldsymbol{0}_{1\times (n-k-1)}
		\end{pmatrix}
		\]
over $\mathbb{F}_{q}$, and constructed several classes of self-dual NMDS codes over $\mathbb{F}_{q}$ with $q$ an odd prime. 
		
		\item In 2021, Liu et al. \cite{LiuTGRSDCC2021} proved that if $k\leq \frac{n-2}{2}$ and $1\leq h\leq k-1$, then the $(\mathcal{L},\mathcal{P})$-TGRS codes with coefficient matrix
\[
		\boldsymbol{B}=
		\begin{pmatrix}
			\boldsymbol{0}_{h\times 1}&\boldsymbol{0}_{h\times n-k-1}\\ 
			b_{h+1,0} &\boldsymbol{0}_{1\times (n-k-1)}\\
			\boldsymbol{0}_{(k-h-1)\times 1}&\boldsymbol{0}_{(k-h-1)\times n-k-1}
		\end{pmatrix}
		\]
		are self-orthogonal.
		
		\item In 2025, Zhang and Ding \cite{ZhangTGRSJAA2025} proved that if $k\geq 4$, then the $(\mathcal{L},\mathcal{P})$-TGRS code with coefficient matrix
\[
		\boldsymbol{B}=
		\begin{pmatrix}
			\boldsymbol{0}_{(k-1)\times 2}&\boldsymbol{0}_{(k-1)\times 1}&\boldsymbol{0}_{(k-1)\times (n-k-3)}\\
			\boldsymbol{0}_{1\times 2}&b_{k-1,2} &\boldsymbol{0}_{1\times (n-k-3)}
		\end{pmatrix}
		\]
		is not self-dual, and further constructed two classes of self-orthogonal $(\mathcal{L},\mathcal{P})$-TGRS codes. 
		
		\item Recently, for general coefficient matrices $\boldsymbol{B}=(b_{i,j})_{k\times (n-k)}$, Hu et al. \cite{HuTGRSTIT2025} presented a sufficient condition for the $(\mathcal{L},\mathcal{P})$-TGRS code to be self-dual and established a necessary and sufficient condition for self-dual $(\mathcal{L},\mathcal{P})$-TGRS codes to be NMDS. Furthermore, they proved that the $(\mathcal{L},\mathcal{P})$-TGRS code with coefficient matrix
\[
		\boldsymbol{B}=
		\begin{pmatrix}
			\boldsymbol{0}_{(k-\ell)\times \ell}&\boldsymbol{0}_{(k-\ell)\times (n-k-\ell)}\\ 
			\boldsymbol{M}&\boldsymbol{0}_{\ell\times (n-k-\ell)}
		\end{pmatrix}
		\]
		is non-GRS for $n\geq 2k$, where
\[
		\boldsymbol{M}=
		\begin{pmatrix}
			b_{k-\ell,0} &0&\cdots&0\\
			b_{k-\ell+1,0} &b_{k-\ell+1,1} &\cdots &0\\
			\vdots &\vdots &\ddots &\vdots\\
			b_{k-1,0} &b_{k-1,1} &\cdots&b_{k-1,\ell-1}
		\end{pmatrix}.
		\]
		{\color{red}Finally,  they further proposed several open problems: characterize the necessary and sufficient conditions under which the $(\mathcal{L},\mathcal{P})$-TGRS code is NMDS in the general case; construct explicit new infinite families of non-GRS MDS codes, NMDS codes, $m$-MDS codes, and self-dual codes from $(\mathcal{L},\mathcal{P})$-TGRS codes ; investigate the dimension of the Schur square of the general $(\mathcal{L},\mathcal{P})$-TGRS code with arbitrary $\boldsymbol{B}$. }
\end{itemize}

Motivated by the above works, in this paper, we consider a special class of the $(\mathcal{L},\mathcal{P})$-TGRS codes with $\ell+1$ twists, and study some coding properties including parity-check matrix, self-orthogonality, NMDS property, LCD MDS property and non-GRS property. {\color{red} More precisely, the main contributions are summarized as follows. Firstly, by discussing in detail the relationship between the rows of the generator matrix and the parity-check matrix for $\mathcal{C}_{k-1}\left(\boldsymbol{\alpha},\boldsymbol{v},\boldsymbol{\eta}\right)$, we determined the existence of self-orthogonal $\mathcal{C}_{k-1}\left(\boldsymbol{\alpha},\boldsymbol{v},\boldsymbol{\eta}\right)$. And then, we obtain some LCD MDS $\mathcal{C}_{k-1}\left(\boldsymbol{\alpha},\boldsymbol{v},\boldsymbol{\eta}\right)$.
	Secondly,  by calculating a special determinant, we give a sufficient and
	necessary  condition for the code $\mathcal{C}_{k-1}\left(\boldsymbol{\alpha},\boldsymbol{v},\boldsymbol{\eta}\right)$ to be NMDS.  Finnaly, by performing some equivalent transformations on a known parity-check matrix of $\mathcal{C}_{k-1}\left(\boldsymbol{\alpha},\boldsymbol{v},\boldsymbol{\eta}\right)$, we get a parity-check matrix of $\mathcal{C}_{k-1}\left(\boldsymbol{\alpha},\boldsymbol{v},\boldsymbol{\eta}\right)$ that has a fairly clear structure, furthermore, by calculating the Schur square dimension of $\mathcal{C}_{k-1}^{\perp}\left(\boldsymbol{\alpha},\boldsymbol{v},\boldsymbol{\eta}\right)$, we prove that $\mathcal{C}_{k-1}\left(\boldsymbol{\alpha},\boldsymbol{v},\boldsymbol{\eta}\right)$ is non-GRS for $2k>n> k+\ell+2$. On the one hand, we discuss the existence of self-orthogonal $\mathcal{C}_{k-1}\left(\boldsymbol{\alpha},\boldsymbol{v},\boldsymbol{\eta}\right)$, which is important for constructing self-dual $\mathcal{C}_{k-1}\left(\boldsymbol{\alpha},\boldsymbol{v},\boldsymbol{\eta}\right)$,  and then constructed LCD MDS $\mathcal{C}_{k-1}\left(\boldsymbol{\alpha},\boldsymbol{v},\boldsymbol{\eta}\right)$; on the other hand, since the twist number $\ell+1$ of $\mathcal{C}_{k-1}\left(\boldsymbol{\alpha},\boldsymbol{v},\boldsymbol{\eta}\right)$  is flexible, but $\mathcal{C}_{k-1}\left(\boldsymbol{\alpha},\boldsymbol{v},\boldsymbol{\eta}\right)$ is just a special case of the general $(\mathcal{L},\mathcal{P})$-TGRS code.  And so, we only partially answered Problems 1-3 proposed by Hu et al. in \cite{HuTGRSTIT2025}.}

This paper is organized as follows. In Section 2, we
introduce some definitions and known results. In Section \ref{sec3}, we obtained a parity-check matrix of this code with a clearer structure. In Section \ref{sec4}, we give some sufficient and necessary conditions, or sufficient conditions for the code $\mathcal{C}_{k-1}\left(\boldsymbol{\alpha},\boldsymbol{v},\boldsymbol{\eta}\right)$ to be self-orthogonal or not, and then give some constructions of LCD MDS codes based on the self-orthogonal  code $\mathcal{C}_{k-1}\left(\boldsymbol{\alpha},\boldsymbol{v},\boldsymbol{\eta}\right)$. In Section \ref{sec5}, we give a sufficient and
necessary  condition for the code $\mathcal{C}_{k-1}\left(\boldsymbol{\alpha},\boldsymbol{v},\boldsymbol{\eta}\right)$ to be NMDS. Finally, we prove that the code $\mathcal{C}_{k-1}\left(\boldsymbol{\alpha},\boldsymbol{v},\boldsymbol{\eta}\right)$ is non-GRS for $2k>n>k+\ell+2$. In Section \ref{sec6}, we give some corresponding examples. In Section \ref{sec7}, we conclude the whole paper. 
\section{Preliminaries}
For convenience, throughout this paper, we consider the code $\mathcal{C}_{k-1}\left(\boldsymbol{\alpha},\boldsymbol{v},\boldsymbol{\eta}\right)$ given in Definition \ref{LPTGRSdefinition} and fix the following notations unless stated otherwise.
\begin{itemize}
\item $q$ is a prime power.
\item $\mathbb{F}_{q}$ is the finite field with $q$ elements.
\item $k$ and $n$ are both positive integers with $2\leq k\leq n.$
\item $\boldsymbol{\alpha}=\left(\alpha_{1}, \ldots, \alpha_{n}\right) \in \mathbb{F}_{q}^{n}$ with $\alpha_{i} \neq \alpha_{j}(i \neq j)$.
\item $\boldsymbol{v}=\left(v_{1}, \ldots, v_{n}\right) \in\left(\mathbb{F}_{q}^{*}\right)^{n}$.
\item $\boldsymbol{\eta}= \left(\eta_{0}, \ldots, \eta_{\ell}\right) \in\mathbb{F}_{q}^{\ell+1}\backslash\left\{\boldsymbol{0}\right\}$ with $0\leq \ell\leq n -k-1$ and $\eta_{\ell}\neq 0$.
\item $u_{i}=\prod\limits_{j=1, j \neq i}^{n}\left(\alpha_{i}-\alpha_{j}\right)^{-1}$ for $1 \leq i \leq n$.  
\item $\dim\left(\mathcal{C}\right)$ is the dimension of a code $\mathcal{C}$.
\item For any positive integers $x,y$ and a sequence $\left\{a_{i}\right\}$, ${\sum\limits^{\prime}}_{i=x}^{y}a_{i}=\begin{cases}
0,&\ \text{if}\ x>y;\\
\sum\limits_{i=x}^{y}a_{i},&\ \text{if}\ x\leq y.
\end{cases}$
\end{itemize} 

In this section, we recall the definitions of the $(\mathcal{L},\mathcal{P})$-twisted generalized Reed-Solomon code and the $t$-th degree complete symmetric polynomial in 
$n$ variables, and then give some necessary lemmas.

The definition of the $(\mathcal{L},\mathcal{P})$-twisted generalized Reed-Solomon code is as follows.
\begin{definition}\label{LPTGRSdefinition}{\rm( \cite{HuTGRSTIT2025}, Definition 2)}
	Let $n$ and $k$ be integers with $0<k\leq n$, $\boldsymbol{B}=(b_{i,j})_{k\times n-k}$, $\mathcal{L}\subseteq\left\{0,1,\ldots,n-k-1\right\}$ and $\mathcal{P}\subseteq\left\{0,1,\ldots,k-1\right\}$. Let  $\boldsymbol{\alpha}=\left(\alpha_{1}, \ldots, \alpha_{n}\right) \in \mathbb{F}_{q}^{n}$ with $\alpha_{i} \neq \alpha_{j}(i \neq j)$, $\boldsymbol{v}=\left(v_{1}, \ldots, v_{n}\right) \in\left(\mathbb{F}_{q}^{*}\right)^{n}$. The $(\mathcal{L},\mathcal{P})$-twisted generalized Reed-Solomon (in short, $(\mathcal{L},\mathcal{P})$-TGRS) code is defined as 
	$$\mathcal{C}\left(\mathcal{L},\mathcal{P},\boldsymbol{B}\right)\triangleq\left\{\left(v_{1}f\left(\alpha_{1}\right), \ldots,v_{n}f\left(\alpha_{n}\right)\right) | f(x) \in \mathcal{F}\left(\mathcal{L},\mathcal{P},\boldsymbol{B}\right)\right\},
	$$
	where $$\mathcal{F}\left(\mathcal{L},\mathcal{P},\boldsymbol{B}\right)=\left\{\sum\limits_{i=0}^{k-1} f_{i} x^{i}+\sum\limits_{i\in\mathcal{P}}f_{i} \sum\limits_{j\in\mathcal{L}}b_{i,j} x^{k+j}|f_{i} \in \mathbb{F}_{q}, 0 \leq i \leq k-1,\right\},$$
	 
\end{definition}

In this paper, we consider a special class of $(\mathcal{L},\mathcal{P})$-TGRS codes, i.e., $$\boldsymbol{B}=\begin{pmatrix}
0&\cdots&0&0&\cdots&0\\
\vdots& &\vdots&\vdots& &\vdots\\
0&\cdots&0&0&\cdots&0\\
\eta_{0}&\cdots&\eta_{\ell}&0&\cdots&0
\end{pmatrix}_{k\times (n-k)}, 0\leq \ell\leq n-k-1.$$ 
And so, by Definition \ref{LPTGRSdefinition}, we know that the corresponding $(\mathcal{L},\mathcal{P})$-TGRS code has the generator matrix 
\begin{equation}\label{+LPTGRSgeneratormattrix}
	\boldsymbol{G}_{k,+}^{(\ell)}=\begin{pmatrix}
		v_{1}&\cdots&v_{n}\\
		v_{1}\alpha_{1}&\cdots&v_{n}\alpha_{n}\\
		\vdots&\ddots&\vdots\\
		v_{1}\alpha_{1}^{k-2}&\cdots&v_{n}\alpha_{n}^{k-2}\\
		v_{1}\left(\alpha_{1}^{k-1}+\sum\limits_{t=0}^{\ell}\eta_{t}\alpha_{1}^{k+t}\right)&\cdots&v_{n}\left(\alpha_{n}^{k-1}+\sum\limits_{t=0}^{\ell}\eta_{t}\alpha_{n}^{k+t}\right)\\
	\end{pmatrix}.
\end{equation}
For convenience, we set  $\boldsymbol{\eta}=\left(\eta_{0}, \ldots, \eta_{\ell}\right) \in\mathbb{F}_{q}^{\ell+1}\backslash\left\{\boldsymbol{0}\right\}$ and briefly denote  the corresponding $(\mathcal{L},\mathcal{P})$-TGRS code as $\mathcal{C}_{k-1}\left(\boldsymbol{\alpha},\boldsymbol{v},\boldsymbol{\eta}\right)$.

The Schur product is defined as follows.
\begin{definition}\label{schurproduct}{\rm( \cite{ZhuTGRSFFA2024}, Definition 2.1)}
For $\boldsymbol{x} = (x_1, \ldots, x_n)$, $\boldsymbol{y} = (y_1, \ldots, y_n) \in \mathbb{F}_q^n$, the Schur product between $\boldsymbol{x}$ and $\boldsymbol{y}$ is defined as
\[
\boldsymbol{x} \star \boldsymbol{y} := (x_1 y_1, \ldots, x_n y_n).
\]
The Schur product of two $q$-ary codes $\mathcal{C}_1$ and $\mathcal{C}_2$ with length $n$ is defined as
\[
\mathcal{C}_1 \star \mathcal{C}_2 = \langle \boldsymbol{c}_1 \star \boldsymbol{c}_2 \mid \boldsymbol{c}_1 \in \mathcal{C}_1, \boldsymbol{c}_2 \in \mathcal{C}_2 \rangle.
\]
Especially, for any code $\mathcal{C}$, $\mathcal{C}^2\triangleq\mathcal{C} \star \mathcal{C}$ is called the Schur square of $\mathcal{C}$.	
\end{definition}

The following Lemma \ref{GRScodeschur2} describes the Schur square of a GRS code and its dual code.
\begin{lemma}\label{GRScodeschur2}{\rm( \cite{ZhuTGRSFFA2024}, Lemma 2.3)}
Let $\boldsymbol{u} = (u_1, \ldots, u_n)$ with $u_j = \prod\limits_{\substack{i=1 \\ i \neq j}}^n (\alpha_j - \alpha_i)^{-1}$ $(j = 1, \ldots, n)$.

$(1)$ If $k \leq \frac{n}{2}$, then $
\mathrm{GRS}_{k,n}(\boldsymbol{\alpha}, \boldsymbol{1}) \star \mathrm{GRS}_{k,n}(\boldsymbol{\alpha}, \boldsymbol{1}) = \mathrm{GRS}_{2k-1,n}(\boldsymbol{\alpha}, \boldsymbol{1});
$ 

$(2)$ if $n\geq k>\frac{n}{2}$, then $
\mathrm{GRS}_{k,n}^\perp(\boldsymbol{\alpha}, \boldsymbol{1}) \star \mathrm{GRS}_{k,n}^\perp(\boldsymbol{\alpha}, \boldsymbol{1}) = \mathrm{GRS}_{2n-2k-1,n}(\boldsymbol{\alpha}, \boldsymbol{u}^2).
$
\end{lemma}
\begin{remark}
By Lemma \ref{GRScodeschur2}, the following two statements are true,

$(1)$ for an $[n,k]$ code $\mathcal{C}$ with $k\leq \frac{n}{2}$, if $\dim\left( \mathcal{C}^2\right)\neq 2k-1$, then $\mathcal{C}$ is non-RS type; 

$(2)$ for an $[n,k]$ code $\mathcal{C}$ with $k> \frac{n}{2}$, if $\dim\left(\left( \mathcal{C}^{\perp}\right)^2\right)\neq 2n-2k-1$, then $\mathcal{C}$ is non-RS type. 
\end{remark}

Next, we recall the definition of the complete symmetric polynomials, as well as a related result. 
\begin{definition}\label{completesymmetricpolynomial} {\rm( \cite{AbdukhalikovnonGRS2026})}
	For any integer $t$, the $t$-th degree complete symmetric polynomial in  $n$ variables is  defined as
	$$S_{t}(x_{1},x_{2},\cdots,x_{n})=\begin{cases}
		0,&\text{if}\ t<0;\\
		\sum\limits_{t_{1}+t_{2}+\cdots+t_{n}=t,t_{i}\geq 0}x_{1}^{t_{1}}x_{2}^{t_{2}}\cdots x_{n}^{t_{n}},&\text{if}\ t\geq 0,\\
	\end{cases}$$
	and denote $S_{t}(x_{1},x_{2},\cdots,x_{n})$ by $S_{t}$.
\end{definition} 
\begin{lemma}\label{usapowersum} {\rm( \cite{AbdukhalikovnonGRS2026}, Lemma 2.4)}
For any subset $\left\{\alpha_{1}, \ldots, \alpha_{n}\right\}\subseteq\mathbb{F}_{q}$ with $n\geq 3$, we have
	$$\sum\limits_{i=1}^{n}u_{i}\alpha_{i}^{h}=\begin{cases}
		0,&\text{if}\ 0\leq h\leq n-2;\\
		S_{h-n+1}(\alpha_{1},\cdots,\alpha_{n}),&\text{if}\ h\geq n-1.\\
	\end{cases}$$ 
\end{lemma}

To give the necessary and sufficient condition for $\mathcal{C}_{k-1}\left(\boldsymbol{\alpha},\boldsymbol{v},\boldsymbol{\eta}\right)$ to be NMDS, the following Lemma \ref{Vandet1} is crucial.
\begin{lemma}\label{Vandet1} {\rm( \cite{SuiTGRSDCC2022}, Lemma 3.2)}
Let $\alpha_1,\ldots,\alpha_{k}$ be distinct elements of $\mathbb{F}_q$, $\mathcal{I}=\left\{1,\ldots,k\right\},$ $\prod\limits_{j\in\mathcal{I}}\left(x-\alpha_{j}\right)=\sum\limits_{j=1}^{k}c_jx^{k-j}$ with  $c_j=0$ for $j>k$, then
$$\det\begin{pmatrix}
	1&\cdots&1\\
	\alpha_{1}&\cdots&\alpha_{k}\\
	\vdots& &\vdots\\
	\alpha_{1}^{h-1}&\cdots&\alpha_{k}^{h-1}\\
	\alpha_{1}^{k+t}&\cdots&\alpha_{k}^{k+t}\\
	\alpha_{1}^{h+1}&\cdots&\alpha_{k}^{h+1}\\
	\vdots& &\vdots\\
	\alpha_{1}^{k-1}&\cdots&\alpha_{k}^{k-1}
\end{pmatrix}=-\boldsymbol{\beta}_{t}\boldsymbol{A}_{\mathcal{I},t}^{-1}\boldsymbol{\gamma}_{t}^{T}\prod\limits_{1\leq j<i\leq k}\left(\alpha_{i}-\alpha_{j}\right),$$
where $\boldsymbol{\beta}_{t}=\left(c_{k+t-h},\ldots,c_{k+1-h},c_{k-h}\right)$, $\boldsymbol{\gamma}_{t}=\left(1,0,\ldots,0\right)\in\mathbb{F}_{q}^{t+1}$ and $$\boldsymbol{A}_{\mathcal{I},t}=\begin{pmatrix}
	1& & & \\
	c_1&1& & \\
	c_2&c_1&1& & \\
	\vdots&\vdots&\ddots&\ddots\\
	c_t&c_{t-1}&\cdots&c_{1}&1\\
\end{pmatrix}.$$
\end{lemma} 

The following Lemmas \ref{MDScondition}-\ref{NMDScondition} provide some necessary and sufficient conditions for a linear
code to be MDS or NMDS, respectively.
\begin{lemma}\label{MDScondition}{\rm( \cite{Huffman2010}, Theorem 2.4.3)}
Let $\mathcal{C}$ be an $[n, k]$ linear code over $\mathbb{F}_{q}$ with $k\geq 1$. Suppose that  $\boldsymbol{G}$ is the 
	generator matrix for $\mathcal{C}$. Hence, $\mathcal{C}$ is MDS if and only if any $k$ columns of $\boldsymbol{G}$ are $\mathbb{F}_{q}$-linearly independent.
\end{lemma}
\begin{lemma}\label{NMDScondition}{\rm( \cite{SuiTGRSTIT2022}, Lemma 3.7)}
Let $\boldsymbol{G}$ be a generator matrix of an $[n,k]$ linear code $\mathcal{C}$. Then $\mathcal{C}$ is NMDS if and only if $\boldsymbol{G}$ satisfies the following conditions simultaneously,

$(1)$ any $k-1$ columns of $\boldsymbol{G}$ are $\mathbb{F}_q$-linearly independent;

$(2)$ there exist $k$ columns of $\boldsymbol{G}$ that are $\mathbb{F}_q$-linearly dependent;

$(3)$ For any $k+1$ columns of $\boldsymbol{G}$, there exist $k$ columns $\mathbb{F}_q$-linearly independent.
\end{lemma}

The following Lemma \ref{selforthogonalMDSLCD} presents a construction method of LCD MDS codes basing on self-orthogonal MDS codes.
\begin{lemma}\label{selforthogonalMDSLCD}{\rm( \cite{LiuTGRSDCC2021}, Lemma 5)}
Let $\mathcal{C}$ be an $\left[n,k,n-k+1\right]$ self-orthogonal MDS linear code generated by the matrix $\boldsymbol{G}=\left[\boldsymbol{A}_{k\times k}:\boldsymbol{B}_{k\times \left(n-k\right)}\right]$. Then for any $\beta\in \mathbb{F}_{q}\backslash\left\{0,-1,1\right\}$,  the linear code generated by the matrix  $\boldsymbol{G}_{\beta}=\left[\boldsymbol{A}_{k\times k}:\beta\boldsymbol{B}_{k\times \left(n-k\right)}\right]$ is an $\left[n,k,n-k+1\right]$ LCD MDS code.
\end{lemma}

The following Lemma \ref{mxroot} is important for constructing the self-orthogonal $(+)$-$(\mathcal{L},\mathcal{P})$-TGRS code.
\begin{lemma}\label{mxroot}{\rm( \cite{LiangLCD*TGRS2026}, Lemma 2.4)}\label{self}
	Let $n\mid (q-1)$, $\lambda\in\mathbb{F}_{q}^{*}$ with $\mathrm{ord}(\lambda)\mid \frac{q-1}{n}$, and $\beta_{1},\ldots,\beta_{n}$ be all roots of $m(x)=x^n-\lambda\in\mathbb{F}_{q}[x]$ in $\mathbb{F}_{q^s}$,  where $s(s\geq 1)$ is an integer. Then $\beta_{i}\in\mathbb{F}_{q}^{*}(1\leq i\leq n)$ and $\beta_{i}\neq \beta_{j}(1\leq i\neq j\leq n)$.
\end{lemma}

\section{The parity-check matrix of the code $\mathcal{C}_{k-1}\left(\alpha,v,\eta\right)$}\label{sec3}  
 
For the code $\mathcal{C}_{k-1}\left(\boldsymbol{\alpha},\boldsymbol{v},\boldsymbol{\eta}\right)$, in order to determine whether  $\mathcal{C}_{k-1}\left(\boldsymbol{\alpha},\boldsymbol{v},\boldsymbol{\eta}\right)$ is self-orthogonal based on the parity-check matrix in Section \ref{sec4}.  In this section, by performing certain elementary transformations for the parity-check matrix given in  \cite{ChengTGRSHTIT2023}, we obtain a more explicit parity-check matrix. 
 
On the one hand, the code $\mathcal{C}_{k-1}\left(\boldsymbol{\alpha},\boldsymbol{v},\boldsymbol{\eta}\right)$ is a special class of $(\mathcal{L},\mathcal{P})$-TGRS codes. When $(\mathcal{L},\mathcal{P})$-TGRS codes were proposed, the parity check matrix of $(\mathcal{L},\mathcal{P})$-TGRS codes with general twists was given in Theorem 3 of the reference  \cite{HuTGRSTIT2025}, while, it is very complicated and so it is difficult to determine whether  $\mathcal{C}_{k-1}\left(\boldsymbol{\alpha},\boldsymbol{v},\boldsymbol{\eta}\right)$ is self-orthogonal by the parity-check matrix.

On the other hand, since the parity-check matrix of a linear code is not unique, for some special classes of $(\mathcal{L},\mathcal{P})$-TGRS codes, scholars have provided some parity-check matrices with other forms. For example, in 2024, for $(\mathcal{L},\mathcal{P})$-TGRS code $\mathcal{C}_{h}\left(\boldsymbol{\alpha},\boldsymbol{v},\boldsymbol{\eta}\right)$ with the generate matrix
$$
\begin{pmatrix}
	v_{1}&\cdots&v_{n}\\
	v_{1}\alpha_{1}&\cdots&v_{n}\alpha_{n}\\
	\vdots&\ddots&\vdots\\
	
	v_{1}\alpha_{1}^{h-1}&\cdots&v_{n}\alpha_{n}^{h-1}\\
	v_{1}\left(\alpha_{1}^{h}+\sum\limits_{i=1}^{\ell^{\prime}}\delta_{i}\alpha_{1}^{k-1+t_{i}}\right)&\cdots&v_{n}\left(\alpha_{n}^{h}+\sum\limits_{i=1}^{\ell^{\prime}}\delta_{i}\alpha_{n}^{k-1+t_{i}}\right)\\
	v_{1}\alpha_{1}^{h+1}&\cdots&v_{n}\alpha_{n}^{h+1}\\
	\vdots&\ddots&\vdots\\
	v_{1}\alpha_{1}^{k-1}&\cdots&v_{n}\alpha_{n}^{k-1}\\
\end{pmatrix},
$$
where $\left(\delta_{1},\ldots,\delta_{\ell^{\prime}}\right)\in\left(\mathbb{F}_{q}^{*}\right)^{\ell^{\prime}}$, $1\leq t_{i}\leq n-k(i=1,\ldots,\ell^{\prime})$ and $1\leq h\leq k-1$, Cheng  \cite{ChengTGRSHTIT2023} gave the following two theorems.
\begin{lemma}\label{CWDparitycheck1}{\rm( \cite{ChengTGRSHTIT2023}, Theorem 5)}
	If $\ell^{\prime}=1$, then $\mathcal{C}_{h}\left(\boldsymbol{\alpha},\boldsymbol{v},\boldsymbol{\delta}\right)$ has the following parity-check matrix
	$$\boldsymbol{H}_{n-k}=\left[\boldsymbol{h}_{1},\ldots,\boldsymbol{h}_{n}\right]$$
	where
	$$
	\boldsymbol{h}_i=\frac{1}{v_{i}}\begin{pmatrix}
		u_{i}\\
		u_{i}\alpha_{i}\\
		\vdots\\
		u_{i}\alpha_{i}^{n-\left(k+t_{\ell^{\prime}}+1\right)}\\
		\psi(u_{i},\alpha_{i})\\
		u_{i}\alpha_{i}^{n-\left(k+t_{\ell^{\prime}}-1\right)}-s_{t_{\ell^{\prime}}}u_{i}\alpha_{i}^{n-\left(k+t_{\ell^{\prime}}\right)}\\
		\vdots\\
		u_{i}\alpha_{i}^{n-\left(k+1\right)}-s_{2}u_{i}\alpha_{i}^{n-\left(k+t_{\ell^{\prime}}\right)}\\
	\end{pmatrix},
	$$
	$$\psi(u_i,\alpha_{i})=\alpha_{i}^{n-(h+1)}-\left(\frac{1+s_{k+t_{\ell^{\prime}}-h}}{\delta_{\ell^{\prime}}}-{\sum^{\prime}}_{f=h+2}^{k}\lambda_{f}s_{k+t_{\ell^{\prime}}-f+1}\right)\alpha_{i}^{n-\left(k+t_{\ell^{\prime}}\right)}-{\sum^{\prime}}_{f=h+2}^{k}\lambda_{f}\alpha_{i}^{n-f},$$ 
	and $s_{y}=\sum\limits_{i=1}^{n}u_{i}\alpha_{i}^{n+y-2}(1\leq y\leq n)$.
\end{lemma}
\begin{remark}\label{ChengYue}
Remark 4  \cite{ChengTGRSHTIT2023} has already shown that Theorem 2.4 (1)  \cite{HuangTGRSDCC2021} is a special case of Lemma \ref{CWDparitycheck1}. 
\end{remark}
\begin{lemma}\label{CWDparitycheck2}{\rm( \cite{ChengTGRSHTIT2023}, Theorem 6)}
If $\ell^{\prime}\geq 2$, then $\mathcal{C}_{h}\left(\boldsymbol{\alpha},\boldsymbol{v},\boldsymbol{\delta}\right)$ has the following parity-check matrix
$$\boldsymbol{H}_{n-k}=\left[\boldsymbol{h}_{1},\ldots,\boldsymbol{h}_{n}\right]$$
where
$$
\boldsymbol{h}_i=\frac{1}{v_{i}}\begin{pmatrix}
	u_{i}\\
	u_{i}\alpha_{i}\\
	\vdots\\
	u_{i}\alpha_{i}^{n-\left(k+t_{\ell^{\prime}}+1\right)}\\
	\psi(u_{i},\alpha_{i})\\
	u_{i}\alpha_{i}^{n-\left(k+t_{\ell^{\prime}}-1\right)}-\frac{a_{h+1,k+t_{\ell^{\prime}}-1}}{\delta_{\ell^{\prime}}}u_{i}\alpha_{i}^{n-\left(k+t_{\ell^{\prime}}\right)}\\
	\vdots\\
	u_{i}\alpha_{i}^{n-\left(k+1\right)}-\frac{a_{h+1,k+1}}{\delta_{\ell^{\prime}}}u_{i}\alpha_{i}^{n-\left(k+t_{\ell^{\prime}}\right)}\\
\end{pmatrix},
$$
$$\psi(u_i,\alpha_{i})=\alpha_{i}^{n-(h+1)}-\left(\frac{a_{h+1,h+1}}{\delta_{\ell^{\prime}}}-{\sum^{\prime}}_{f=h+2}^{k}\lambda_{f}\frac{a_{h+1,f}}{\delta_{\ell^{\prime}}}\right)\alpha_{i}^{n-\left(k+t_{\ell^{\prime}}\right)}-{\sum^{\prime}}_{f=h+2}^{k}\lambda_{f}\alpha_{i}^{n-f},$$
$$
a_{h+1,j}=\begin{cases}
	\sum\limits_{r=1}^{{\ell^{\prime}}}\delta_{r}s_{k+t_r-j+1}+s_{h+1-j+1},&1\leq j\leq h+1;\\
	\sum\limits_{r=1}^{{\ell^{\prime}}}\delta_{r}s_{k+t_r-j+1},&h+2\leq j\leq k+t_{1};\\
	\sum\limits_{r=i}^{{\ell^{\prime}}}\delta_{r}s_{k+t_r-j+1},&k+t_{i-1}\leq j\leq k+t_{i}\ \text{for}\ 2\leq i\leq {\ell^{\prime}};\\
	0,&k+t_{{\ell^{\prime}}}+1\leq j\leq n;
\end{cases}
$$
and $s_{y}=\sum\limits_{i=1}^{n}u_{i}\alpha_{i}^{n+y-2}(1\leq y\leq n)$. 
\end{lemma}
\begin{remark}
The author  \cite{ChengTGRSHTIT2023} proved  Lemma \ref{CWDparitycheck2}, by only using the condition $\delta_{\ell^{\prime}}\in \mathbb{F}_{q}^{*}$ for the vector $\left(\delta_{1},\ldots,\delta_{\ell^{\prime}}\right)$, and so we can assume that $\left(\delta_{1},\ldots,\delta_{\ell^{\prime}}\right)\in\mathbb{F}_{q}^{\ell^{\prime}}\backslash\left\{\boldsymbol{0}\right\}$.
\end{remark}

By taking $h=k-1$ and $\ell^{\prime}=\ell+1$, $\left(t_1,t_2,\ldots,t_{\ell^{\prime}}\right)=\left(1,2,\ldots,\ell+1\right)$, and $\delta_{i}=\eta_{i-1}(i=1,2,\ldots,\ell+1)$ in Lemmas  \ref{CWDparitycheck1}-\ref{CWDparitycheck2}, we can obtain a parity-check maritx of the code $\mathcal{C}_{k-1}\left(\boldsymbol{\alpha},\boldsymbol{v},\boldsymbol{\eta}\right)$. Note that the expression of $a_{h+1,j}$ is complex in Lemma \ref{CWDparitycheck2}. For the parity-check matrix in Lemma \ref{CWDparitycheck2}, we can obtain the following theorem through certain elementary transformations.
\begin{theorem}\label{+LPTGRSparitymatrix}
	Let $$\varTheta_{i}=\frac{\sum\limits_{t=0}^{\ell}\eta_{t}S_{k+t+i-n+1}}{1+\sum\limits_{t=0}^{\ell}\eta_{t}S_{t+1}}(n-k-\ell-1\leq i\leq n-k-1)$$ 
	and
	$$\varOmega_{i}=\frac{\sum\limits_{t=0}^{\ell}\eta_{t}S_{k+t+i-n+1}}{\eta_{\ell}}(n-k-\ell\leq i\leq n-k-1).$$
	Then the following two statements are true. 
	
	$(1)$ If $1+\sum\limits_{t=0}^{\ell}\eta_{t}S_{t+1}\neq 0$, then the matrix 
	\begin{equation}\label{paritycheckmatrix1}
		\boldsymbol{H}_{n-k,+,1}^{(\ell,n)}=\begin{pmatrix}
			\cdots&\frac{u_{j}}{v_{j}}&\cdots\\
			\cdots&\frac{u_{j}}{v_{j}}\alpha_{j}&\cdots\\
			\vdots&\vdots&\vdots\\
			\cdots&\frac{u_{j}}{v_{j}}\alpha_{j}^{n-k-(\ell+2)}&\cdots\\
			\cdots&\frac{u_{j}}{v_{j}}\left(\alpha_{j}^{n-k-(\ell+1)}-\varTheta_{n-k-(\ell+1)}\alpha_{j}^{n-k}\right)&\cdots\\ 
			\vdots&\vdots&\vdots\\
			\cdots&\frac{u_{j}}{v_{j}}\left(\alpha_{j}^{n-k-1}-\varTheta_{n-k-1}\alpha_{j}^{n-k}\right)&\cdots
		\end{pmatrix}_{(n-k)\times n}
	\end{equation}
	is a parity-check matrix of the code $\mathcal{C}_{k-1}\left(\boldsymbol{\alpha},\boldsymbol{v},\boldsymbol{\eta}\right)$. 
	
	$(2)$ If $\ell\geq 1$ and  $1+\sum\limits_{t=0}^{\ell}\eta_{t}S_{t+1}=0$, then the matrix
	\begin{equation}\label{paritymatrix2}
		\boldsymbol{H}_{n-k,+,2}^{(\ell,n)}=\begin{pmatrix}
			\cdots&\frac{u_{j}}{v_{j}}&\cdots\\
			\cdots&\frac{u_{j}}{v_{j}}\alpha_{j}&\cdots\\
			\vdots&\vdots&\vdots\\
			\cdots&\frac{u_{j}}{v_{j}}\alpha_{j}^{n-k-(\ell+2)}&\cdots\\
			\cdots&\frac{u_{j}}{v_{j}}\left(\alpha_{j}^{n-k-\ell}-\varOmega_{n-k-\ell}\alpha_{j}^{n-k-(\ell+1)}\right)&\cdots\\
			\vdots&\vdots&\vdots\\
			\cdots&\frac{u_{j}}{v_{j}}\left(\alpha_{j}^{n-k-1}-\varOmega_{n-k-1}\alpha_{j}^{n-k-(\ell+1)}\right)&\cdots\\
			\cdots&\frac{u_{j}}{v_{j}}\alpha_{j}^{n-k}&\cdots
		\end{pmatrix}_{(n-k)\times n}
	\end{equation}
	is a parity-check matrix of the code $\mathcal{C}_{k-1}\left(\boldsymbol{\alpha},\boldsymbol{v},\boldsymbol{\eta}\right)$. 
\end{theorem}
{\bf Proof.} Firstly, by taking $\ell=0$ and $1+\sum\limits_{t=0}^{\ell}\eta_{t}S_{t+1}\neq0$ in Theorem \ref{+LPTGRSparitymatrix}, the corresponding result is just Theorem 2.4 (1)  \cite{HuangTGRSDCC2021}. And so, by Remark \ref{ChengYue}, when $\ell=0$ and $1+\sum\limits_{t=0}^{\ell}\eta_{t}S_{t+1}\neq0$, Theorem \ref{+LPTGRSparitymatrix} is just a special case of Lemma \ref{CWDparitycheck1}.

Next, for the case $\ell\geq 1$. By taking $h=k-1$ and $\ell^{\prime}=\ell+1$, $\left(t_1,t_2,\ldots,t_{\ell^{\prime}}\right)=\left(1,2,\ldots,\ell+1\right)$, and $\delta_{i}=\eta_{i-1}(i=1,2,\ldots,\ell+1)$ in Lemma \ref{CWDparitycheck2}, we can get   
$$
\boldsymbol{h}_i=\frac{u_i}{v_i}\begin{pmatrix}
	1\\
	\alpha_{i}\\
	\vdots\\
	\alpha_{i}^{n-\left(k+\ell+2\right)}\\
	\alpha_{i}^{n-k}-\frac{\sum\limits_{r=1}^{{\ell+1}}\delta_{r}s_{r+1}+1}{\eta_{\ell}}\alpha_{i}^{n-\left(k+\ell+1\right)}\\
	\alpha_{i}^{n-\left(k+\ell\right)}-\frac{a_{k,k+\ell}}{\eta_{\ell}}\alpha_{i}^{n-\left(k+\ell+1\right)}\\
	\vdots\\
	\alpha_{i}^{n-\left(k+1\right)}-\frac{a_{k,k+1}}{\eta_{\ell}}\alpha_{i}^{n-\left(k+\ell+1\right)}\\
\end{pmatrix}.
$$
For $\ell\geq 1$ and $1+\sum\limits_{t=0}^{\ell}\eta_{t}S_{t+1}=0$, by directly calculating, it is easy to prove that $\sum\limits_{r=1}^{{\ell+1}}\delta_{r}s_{r+1}+1=0$ when $1+\sum\limits_{t=0}^{\ell}\eta_{t}S_{t+1}=0$ and $\frac{a_{k,j}}{\eta_{\ell}}=\varOmega_{n-j}(j=1,\ldots,k+\ell)$, and so, Theorem \ref{+LPTGRSparitymatrix} (2) is just a special case of Lemma \ref{CWDparitycheck2}.
 
For $\ell\geq 1$ and $1+\sum\limits_{t=0}^{\ell}\eta_{t}S_{t+1}\neq0$, it is easy to know that $$\boldsymbol{H}_{n-k}=\mathrm{diag}\left\{1,\ldots,1,\frac{\sum\limits_{r=1}^{{\ell+1}}\delta_{r}s_{r+1}+1}{\eta_{\ell}},1,\ldots,1\right\}\widetilde{\boldsymbol{H}_{n-k}}$$
where $\widetilde{\boldsymbol{H}_{n-k}}=\left[\widetilde{\boldsymbol{h}_1},\ldots,\widetilde{\boldsymbol{h}_{n}}\right]$, and
$$
\boldsymbol{h}_i=\mathrm{diag}\left\{1,\ldots,1,\frac{\sum\limits_{r=1}^{{\ell+1}}\delta_{r}s_{r+1}+1}{\eta_{\ell}},1,\ldots,1\right\}\widetilde{\boldsymbol{h}_i}$$
with $$\widetilde{\boldsymbol{h}_i}=\frac{u_i}{v_i}\begin{pmatrix}
	1\\
	\alpha_{i}\\
	\vdots\\
	\alpha_{i}^{n-\left(k+\ell+2\right)}\\
	\frac{\eta_{\ell}}{\sum\limits_{r=1}^{{\ell+1}}\delta_{r}s_{r+1}+1}\alpha_{i}^{n-k}-\alpha_{i}^{n-\left(k+\ell+1\right)}\\
	\alpha_{i}^{n-\left(k+\ell\right)}-\frac{a_{k,k+\ell}}{\eta_{\ell}}\alpha_{i}^{n-\left(k+\ell+1\right)}\\
	\vdots\\
	\alpha_{i}^{n-\left(k+1\right)}-\frac{a_{k,k+1}}{\eta_{\ell}}\alpha_{i}^{n-\left(k+\ell+1\right)}\\
\end{pmatrix}.
$$
By $1+\sum\limits_{t=0}^{\ell}\eta_{t}S_{t+1}\neq0$, we have $\sum\limits_{r=1}^{{\ell+1}}\delta_{r}s_{r+1}+1\neq 0$. Furthermore, since $$\mathrm{diag}\left\{1,\ldots,1,\frac{\sum\limits_{r=1}^{{\ell+1}}\delta_{r}s_{r+1}+1}{\eta_{\ell}},1,\ldots,1\right\}$$ is non-singular over $\mathbb{F}_{q}$, and so $\widetilde{\boldsymbol{H}_{n-k}}$ is a parity-check matrix of the code $\mathcal{C}_{k-1}\left(\boldsymbol{\alpha},\boldsymbol{v},\boldsymbol{\eta}\right)$. Now, by replacing the $j$-th row of $\widetilde{\boldsymbol{h}_i}$ with $\alpha_{i}^{j-1}-\frac{a_{k,n-j+1}}{\sum\limits_{r=1}^{{\ell+1}}\delta_{r}s_{r+1}+1}\alpha_{i}^{n-k}$ for each $j=n-(k+\ell-1),\ldots,n-k$, and directly calculating, it is easy to prove that $\frac{a_{k,n-j+1}}{\sum\limits_{r=1}^{{\ell+1}}\delta_{r}s_{r+1}+1}=\varTheta_{j-1}$ for each $j=n-(k+\ell-1),\ldots,n-k$, and so, Theorem \ref{+LPTGRSparitymatrix} (1) is just a special case of Lemma \ref{CWDparitycheck2}.
 
\begin{remark}
By taking $\ell=0$ and $\boldsymbol{\eta}=\eta_{0}\in\mathbb{F}_{q}^{*}$, or $\ell=1$ and $\boldsymbol{\eta}=(0,\eta_{1})\in\mathbb{F}_{q}^{2}\backslash\left\{\boldsymbol{0}\right\}$ in Theorem \ref{+LPTGRSparitymatrix}, the corresponding results are just Theorem 2.4 $(1)$ in  \cite{HuangTGRSDCC2021} and Theorem 4.1 in  \cite{DingTGRSAMC2025},  respectively.
\end{remark} 
\section{The existence for self-orthogonal $\mathcal{C}\left(\alpha,v,\eta\right)$ and their application to LCD MDS}\label{sec4}
In this section, for the code $\mathcal{C}_{k-1}\left(\boldsymbol{\alpha},\boldsymbol{v},\boldsymbol{\eta}\right)$, by analyzing the inclusion relationship between the code $\mathcal{C}_{k-1}\left(\boldsymbol{\alpha},\boldsymbol{v},\boldsymbol{\eta}\right)$ and its dual code, we give some sufficient and necessary  conditions and sufficient conditions for the code  $\mathcal{C}_{k-1}\left(\boldsymbol{\alpha},\boldsymbol{v},\boldsymbol{\eta}\right)$ to be self-orthogonal  or not.
\subsection{The case $\ell=0$}
In this subsection, we discuss the existence for the self-orthogonal code $\mathcal{C}_{k-1}\left(\boldsymbol{\alpha},\boldsymbol{v},\boldsymbol{\eta}\right)$ when $\ell=0$.

 Firstly, four sufficient conditions for the code $\mathcal{C}_{k-1}\left(\boldsymbol{\alpha},\boldsymbol{v},\boldsymbol{\eta}\right)$ not to be self-orthogonal are given as follows.
\begin{theorem}\label{l=0not}
	If one of the following conditions is satisfied, then the code $\mathcal{C}_{k-1}\left(\boldsymbol{\alpha},\boldsymbol{v},\boldsymbol{\eta}\right)$ is not self-orthogonal.
	
	$(1)$ $n=2k\geq 6$, $\sum\limits_{i=1}^{n}\alpha_{i}\neq 0$ and  $1+\eta_{0}\sum\limits_{i=1}^{n}\alpha_{i}= 0$; 
	
	$(2)$  $n=2k\geq 6$, $\sum\limits_{i=1}^{n}\alpha_{i}\neq 0$, $1+\eta_{0}\sum\limits_{i=1}^{n}\alpha_{i}\neq 0$ and  $\mathrm{Char}\left(\mathbb{F}_{q}\right)=2$;
	 
	$(3)$ $n=2k\geq 6$, $\sum\limits_{i=1}^{n}\alpha_{i}=0$ and  $\mathrm{Char}\left(\mathbb{F}_{q}\right)\neq 2$;
	
	$(4)$ $n=2k+1$ and there exists \(\lambda\in\mathbb F_q^*\) such that $
	v_i^2=\lambda u_i$ for $1\le i\le n$. 
\end{theorem}
{\bf Proof.} By $(\ref{+LPTGRSgeneratormattrix})$, we know that the code $\mathcal{C}_{k-1}\left(\boldsymbol{\alpha},\boldsymbol{v},\boldsymbol{\eta}\right)$ with $\ell=0$ has the generator matrix 
$$\boldsymbol{G}_{k,+}^{(0)}=\begin{pmatrix}
	v_{1}&\cdots&v_{n}\\
	v_{1}\alpha_{1}&\cdots&v_{n}\alpha_{n}\\
	\vdots& &\vdots\\
	v_{1}\alpha_{1}^{k-2}&\cdots&v_{n}\alpha_{n}^{k-2}\\
	v_{1}\left(\alpha_{1}^{k-1}+\eta_{0}\alpha_{1}^{k}\right)&\cdots&v_{n}\left(\alpha_{n}^{k-1}+\eta_{0}\alpha_{n}^{k}\right)\\
\end{pmatrix}=\begin{pmatrix}
	\boldsymbol{g}_{0}^{(0)}\\
	\boldsymbol{g}_{1}^{(0)}\\
	\vdots\\
	\boldsymbol{g}_{k-2}^{(0)}\\
	\boldsymbol{g}_{k-1}^{(0)}
\end{pmatrix}.$$

{\bf (1)} By Theorem 2.4 (3) of the reference  \cite{HuangTGRSDCC2021}, for $\ell=0$, $\sum\limits_{i=1}^{n}\alpha_{i}\neq 0$, $1+\eta_{0}\sum\limits_{i=1}^{n}\alpha_{i}=0$ and $n=2k$, the code $\mathcal{C}_{k-1}\left(\boldsymbol{\alpha},\boldsymbol{v},\boldsymbol{\eta}\right)$ has the parity-check matrix  
$$
\boldsymbol{H}_{k,+,3}^{(0,2k)}=\begin{pmatrix}
	\frac{u_{1}}{v_{1}}&\cdots&\frac{u_{n}}{v_{n}}\\
	\frac{u_{1}}{v_{1}}\alpha_{1}&\cdots&\frac{u_{n}}{v_{n}}\alpha_{n}\\
	\vdots& &\vdots\\
	\frac{u_{1}}{v_{1}}\alpha_{1}^{k-2}&\cdots&\frac{u_{n}}{v_{n}}\alpha_{n}^{k-2}\\
	\frac{u_{1}}{v_{1}}\alpha_{1}^{k}&\cdots&\frac{u_{n}}{v_{n}}\alpha_{n}^{k}
\end{pmatrix}=\begin{pmatrix}
	\boldsymbol{h}_{0,3}^{(0,2k)}\\
	\vdots\\
	\boldsymbol{h}_{k-2,3}^{(0,2k)}\\ 
	\boldsymbol{h}_{k,3}^{(0,2k)}
\end{pmatrix}.
$$
If
\(\mathcal C_{k-1}(\boldsymbol{\alpha},\boldsymbol{v},\boldsymbol{\eta})\)
is self-orthogonal, then $\boldsymbol{G}_{k,+}^{(0)}\left(\boldsymbol{G}_{k,+}^{(0)}\right)^T=0$, it means $M_t=0$ for $0\le t\le2k-4,$ and $M_{k-1+s}+\eta_0M_{k+s}=0$ for $0\le s\le k-2,$
where $w_i=v_i^2$ and $M_t=\sum\limits_{i=1}^{n}w_i\alpha_i^t.$
Since \(k\ge3\), for \(0\le s\le k-3\) we have
$M_{k-1+s}=0$, and then $M_{k+s}=0$. Thus $M_k=M_{k+1}=\cdots=M_{2k-3}=0.$ Thus $M_{2k-2}=0.$
Therefore
\[
M_0=M_1=\cdots=M_{2k-2}=0.
\]
It means that $w=(v_1^2,\ldots,v_n^2)\neq 0$ is a nonzero solution of the homogeneous system $\boldsymbol{V}_{(2k-1)\times2k}\boldsymbol{x}=\boldsymbol{0}$, where $
\boldsymbol{V}_{(2k-1)\times2k}=\bigl(\alpha_i^t\bigr)_{0\le t\le2k-2,\ 1\le i\le n}$. Since the
\(\alpha_i\) are distinct, $\rank\left(\boldsymbol{V}_{(2k-1)\times2k}\right)=2k-1$. Hence $\dim\ker(\boldsymbol{V}_{(2k-1)\times2k})=1.$ Note that $\sum\limits_{i=1}^{n}u_i\alpha_i^t=0$ for $0\le t\le2k-2,$
so $$\boldsymbol{0}\neq\boldsymbol{u}=(u_1,\ldots,u_n)\in\ker\left(\boldsymbol{V}_{(2k-1)\times2k}\right).$$ Furthermore, $\ker(\boldsymbol{V}_{(2k-1)\times2k})=\operatorname{span}(\boldsymbol{u}).$ Thus there exists \(\lambda\in\mathbb F_q^*\) such that $
v_i^2=\lambda u_i$ for $1\le i\le n.$ 

By the suppose, we know $g_{k-1}^{(0)}\in
\mathcal C_{k-1}^{\perp}(\boldsymbol{\alpha},\boldsymbol{v},\boldsymbol{\eta}).$
Then there exist \(b_0,b_1,\ldots,b_{k-2},b_k\in\mathbb F_q\) such that for every \(1\le i\le n\),
\[
v_i(\alpha_i^{k-1}+\eta_0\alpha_i^k)
=
\frac{u_i}{v_i}
\left(
b_0+b_1\alpha_i+\cdots
+b_{k-2}\alpha_i^{k-2}
+b_k\alpha_i^k
\right),
\] 
it means that the polynomials
$$
\lambda(x^{k-1}+\eta_0x^k)-\left(b_0+b_1x+\cdots+b_{k-2}x^{k-2}+b_kx^k\right)$$
has $n$ distinct roots, which is contradict with $n=2k> k$, i.e., $$\lambda(x^{k-1}+\eta_0x^k)=\left(b_0+b_1x+\cdots+b_{k-2}x^{k-2}+b_kx^k\right).$$ By comparing the coefficient of \(x^{k-1}\) yields
\(\lambda=0\), a contradiction. Therefore
\[
g_{k-1}^{(0)}\notin
\mathcal C_{k-1}^{\perp}(\boldsymbol{\alpha},\boldsymbol{v},\boldsymbol{\eta}),
\]
so the code is not self-orthogonal.

{\bf (2)}  By Theorem 2.4 (1) of the reference  \cite{HuangTGRSDCC2021}, for $\ell=0$, $\sum\limits_{i=1}^{n}\alpha_{i}\neq 0$, $1+\eta_{0}\sum\limits_{i=1}^{n}\alpha_{i}\neq 0$ and $n=2k$, the code $\mathcal{C}_{k-1}\left(\boldsymbol{\alpha},\boldsymbol{v},\boldsymbol{\eta}\right)$ has the parity-check matrix  
$$
\boldsymbol{H}_{k,+,1}^{(0,2k)}=\begin{pmatrix}
	\frac{u_{1}}{v_{1}}&\cdots&\frac{u_{n}}{v_{n}}\\
	\frac{u_{1}}{v_{1}}\alpha_{1}&\cdots&\frac{u_{n}}{v_{n}}\alpha_{n}\\
	\vdots& &\vdots\\
	\frac{u_{1}}{v_{1}}\alpha_{1}^{k-2}&\cdots&\frac{u_{n}}{v_{n}}\alpha_{n}^{k-2}\\
	\frac{u_{1}}{v_{1}}\left(\alpha_{1}^{k-1}-\frac{\eta_{0}}{1+\eta_{0}\sum\limits_{i=1}^{n}\alpha_{i}}\alpha_{1}^{k}\right)&\cdots&\frac{u_{n}}{v_{n}}\left(\alpha_{n}^{k-1}-\frac{\eta_{0}}{1+\eta_{0}\sum\limits_{i=1}^{n}\alpha_{i}}\alpha_{n}^{k}\right)
\end{pmatrix}=\begin{pmatrix}
	\boldsymbol{h}_{0,1}^{(0,2k)}\\
	\vdots\\
	\boldsymbol{h}_{k-2,1}^{(0,2k)}\\ 
	\boldsymbol{h}_{k-1,1}^{(0,2k)}
\end{pmatrix}.
$$

To prove our results, we use the method of contradiction, i.e., we assume that the code $\mathcal{C}_{k-1}\left(\boldsymbol{\alpha},\boldsymbol{v},\boldsymbol{\eta}\right)$ is self-orthogonal, then we will get a contradiction.

Now we assume that the code $\mathcal{C}_{k-1}\left(\boldsymbol{\alpha},\boldsymbol{v},\boldsymbol{\eta}\right)$ is self-orthogonal, then $\boldsymbol{G}_{k,+}^{(0)}\left(\boldsymbol{G}_{k,+}^{(0)}\right)^T=0$ and 
\begin{equation}\label{th4.1(1)}
	\boldsymbol{g}_{k-1}^{(0)}\in\mathrm{Span}_{\mathbb{F}_{q}}\left\{\boldsymbol{h}_{0,1}^{(0,2k)},\ldots,\boldsymbol{h}_{k-2,1}^{(0,2k)},\boldsymbol{h}_{k-1,1}^{(0,2k)}\right\}.
\end{equation}
By $\boldsymbol{G}_{k,+}^{(0)}\left(\boldsymbol{G}_{k,+}^{(0)}\right)^T=0$, we know that there exists \(\lambda\in\mathbb F_q^*\) such that $
v_i^2=\lambda u_i$ for $1\le i\le n.$ By \eqref{th4.1(1)}, we know that for any $1\leq i\leq n$, there exists some $a_{j}(0\leq j\leq k-1)$ not all zero such that 
$$v_{i}\left(\alpha_{i}^{k-1}+\eta_{0}\alpha_{i}^{k}\right)=\frac{u_{i}}{v_{i}}\left(a_{0}+a_{1}\alpha_{i}+\cdots+a_{k-2}\alpha_{i}^{k-2}+a_{k-1}\left(\alpha_{i}^{k-1}-\frac{\eta_{0}}{1+\eta_{0}\sum\limits_{s=1}^{n}\alpha_{s}}\alpha_{i}^{k}\right)\right),$$
i.e., the polynomial $$P(x)=\lambda\left(x^{k-1}+\eta_{0}x^{k}\right)-\left(a_{0}+a_{1}x+\cdots+a_{k-2}x^{k-2}+a_{k-1}\left(x^{k-1}-\frac{\eta_{0}}{1+\eta_{0}\sum\limits_{s=1}^{n}\alpha_{s}}x^{k}\right)\right)$$ has $n$ distinct roots. By $n=2k>k$, we know $P(x)=0$, i.e., $$\begin{cases}
	a_{j}=0,&\text{for} \ 0\leq j\leq k-2;\\
	a_{k-1}=\lambda;& \\
	a_{k-1}\cdot\frac{-\eta_{0}}{1+\eta_{0}\sum\limits_{s=1}^{n}\alpha_{s}}=\lambda\eta_{0}.& \\
\end{cases}$$
Now by $a_{k-1}=\lambda$, we have 
$$\frac{-\lambda\eta_{0}}{1+\eta_{0}\sum\limits_{s=1}^{n}\alpha_{s}}=\lambda\eta_{0},$$
furthermore, by $\eta_{0}, \lambda\in\mathbb{F}_{q}^{*}$, $1+\eta_{0}\sum\limits_{s=1}^{n}\alpha_{s}\neq 0$ and $\mathrm{Char}\left(\mathbb{F}_{q}\right)=2$, we have
$$\sum\limits_{s=1}^{n}\alpha_{s}=0,$$
it's a contradiction. Then the code $\mathcal{C}_{k-1}\left(\boldsymbol{\alpha},\boldsymbol{v},\boldsymbol{\eta}\right)$ is not self-orthogonal.

{\bf (3)} In the  same proof  as that of the above (1), if $\mathcal{C}_{k-1}\left(\boldsymbol{\alpha},\boldsymbol{v},\boldsymbol{\eta}\right)$ is self-orthogonal, then there exists \(\lambda\in\mathbb F_q^*\) such that $
v_i^2=\lambda u_i$ for $1\le i\le n,$ furthermore, we can get $
-\lambda\eta_{0}=\lambda\eta_{0},$ i.e., $2=0$. By $\mathrm{Char}\left(\mathbb{F}_{q}\right)\neq 2$, it is a contradiction, then the code $\mathcal{C}_{k-1}\left(\boldsymbol{\alpha},\boldsymbol{v},\boldsymbol{\eta}\right)$ is not self-orthogonal.

{\bf (4)} For $\sum\limits_{i=1}^{n}\alpha_{i}\neq 0$ and $1+\eta_{0}\sum\limits_{i=1}^{n}\alpha_{i}\neq 0$, since there exists \(\lambda\in\mathbb F_q^*\) such that $
v_i^2=\lambda u_i$ for $1\le i\le n$, in the same proof  as that of Theorem \ref{l=0not} (2), we can get
$$\frac{-v_{i}\eta_{0}^{2}}{1+\eta_{0}\sum\limits_{s=1}^{n}\alpha_{s}}=0.$$
By $\eta_{0}, v_{i}\in\mathbb{F}_{q}^{*}$, it is a contradiction. So the code $\mathcal{C}_{k-1}\left(\boldsymbol{\alpha},\boldsymbol{v},\boldsymbol{\eta}\right)$ is not self-orthogonal.

For $\sum\limits_{i=1}^{n}\alpha_{i}=0$. In the  same proof  as that of Theorem \ref{l=0not} (2), we can get $v_{i}\eta_{0}^{2}=0,$ which is a contradiction. Then the code $\mathcal{C}_{k-1}\left(\boldsymbol{\alpha},\boldsymbol{v},\boldsymbol{\eta}\right)$ is not self-orthogonal.

For $\sum\limits_{i=1}^{n}\alpha_{i}\neq 0$ and $1+\eta_{0}\sum\limits_{i=1}^{n}\alpha_{i}=0$. Assume that $\mathcal{C}_{k-1}\left(\boldsymbol{\alpha},\boldsymbol{v},\boldsymbol{\eta}\right)$ is self-orthogonal, hence, $g_{k-1}^{(0)}\in\mathcal C_{k-1}^{\perp}(\boldsymbol{\alpha},\boldsymbol{v},\boldsymbol{\eta}),$ we know that there exist \(d_0,d_1,\ldots,d_{k-1},d_{k+1}\in\mathbb F_q\) such that for every \(1\le i\le n\), 
$$v_{i}\left(\alpha_{i}^{k-1}+\eta_{0}\alpha_{i}^{k}\right)=\frac{u_{i}}{v_{i}}\left(d_{0}+d_{1}\alpha_{i}+\cdots+d_{k-1}\alpha_{i}^{k-1}+d_{k+1}\alpha_{i}^{k+1}\right),$$ 
since there exists \(\lambda\in\mathbb F_q^*\) such that $
v_i^2=\lambda u_i$ for $1\le i\le n,$ it means that the polynomials
$$
\lambda(x^{k-1}+\eta_0x^k)-\left(d_0+d_1x+\cdots+d_{k-1}x^{k-1}+d_{k+1}x^{k+1}\right)$$
has $n$ distinct roots, which is contradict with $n=2k+1> k+1$, i.e., $$\lambda(x^{k-1}+\eta_0x^k)=\left(d_0+d_1x+\cdots+d_{k-1}x^{k-1}+d_{k+1}x^{k+1}\right).$$ By comparing the coefficient of \(x^{k}\) yields
\(\lambda=0\), a contradiction. Therefore
\[
g_{k-1}^{(0)}\notin
\mathcal C_{k-1}^{\perp}(\boldsymbol{\alpha},\boldsymbol{v},\boldsymbol{\eta}),
\]
so the code is not self-orthogonal.  

From the above discussions, we complete the proof of Theorem \ref{l=0not}. 

$\hfill\Box$

Next,  we give a sufficient condition for the code $\mathcal{C}_{k-1}\left(\boldsymbol{\alpha},\boldsymbol{v},\boldsymbol{\eta}\right)$ to be self-orthogonal.
\begin{theorem}\label{l=0yes}
	If $\ell=0$, $n=2k$ with $k\geq 2$, $\sum\limits_{i=1}^{n}\alpha_{i}=0$, $\mathrm{Char}\left(\mathbb{F}_{q}\right)=2$, and there exists some  $\lambda\in\mathbb{F}_{q}^{*}$ such that $\lambda u_{i}=v_{i}^{2}$ for $1 \leq i \leq n$, then the code $\mathcal{C}_{k-1}\left(\boldsymbol{\alpha},\boldsymbol{v},\boldsymbol{\eta}\right)$ is self-orthogonal. 
\end{theorem}
{\bf Proof}. By Theorem 2.4 (2) of the reference  \cite{HuangTGRSDCC2021}, for $\ell=0$, $n=2k$ and $\sum\limits_{i=1}^{n}\alpha_{i}=0$,the code $\mathcal{C}_{k-1}\left(\boldsymbol{\alpha},\boldsymbol{v},\boldsymbol{\eta}\right)$ has the parity-check matrix  
$$
\boldsymbol{H}_{k,+,2}^{(0,2k)}=\begin{pmatrix}
	\frac{u_{1}}{v_{1}}&\cdots&\frac{u_{n}}{v_{n}}\\
	\frac{u_{1}}{v_{1}}\alpha_{1}&\cdots&\frac{u_{n}}{v_{n}}\alpha_{n}\\
	\vdots& &\vdots\\
	\frac{u_{1}}{v_{1}}\alpha_{1}^{k-2}&\cdots&\frac{u_{n}}{v_{n}}\alpha_{n}^{k-2}\\
	\frac{u_{1}}{v_{1}}\left(\alpha_{1}^{k-1}-\eta_{0}\alpha_{1}^{k}\right)&\cdots&\frac{u_{n}}{v_{n}}\left(\alpha_{n}^{k-1}-\eta_{0}\alpha_{n}^{k}\right)
\end{pmatrix}=\begin{pmatrix}
	\boldsymbol{h}_{0,2}^{(0,2k)}\\
	\vdots\\
	\boldsymbol{h}_{k-2,2}^{(0,2k)}\\ 
	\boldsymbol{h}_{k-1,2}^{(0,2k)}
\end{pmatrix}.
$$
It's easy to know that $\boldsymbol{g}_{i}^{(0)}(0\leq i\leq k-2)$ can be represented by $\boldsymbol{h}_{0,2}^{(0,2k)},\ldots,\boldsymbol{h}_{k-2,2}^{(0,2k)}$, i.e., $$\boldsymbol{g}_{i}^{(0)}\in\mathcal{C}_{k-1}^{\perp}\left(\boldsymbol{\alpha},\boldsymbol{v},\boldsymbol{\eta}\right)(0\leq i\leq k-2).$$ Furthermore, we only need to prove that $\boldsymbol{g}_{k-1}^{(0)}\in\mathcal{C}_{k-1}^{\perp}\left(\boldsymbol{\alpha},\boldsymbol{v},\boldsymbol{\eta}\right).$ In fact, by $\lambda u_{i}=v_{i}^{2}(1 \leq i \leq n)$ and $\mathrm{Char}\left(\mathbb{F}_{q}\right)=2$, we know that there exist $$o_{i}=0(0\leq i\leq k-2),o_{k-1}=\lambda,$$ such that 
$$v_{i}\left(\alpha_{i}^{k-1}+\eta_{0}\alpha_{i}^{k}\right)=\frac{u_{i}}{v_{i}}\left(o_{0}+o_{1}\alpha_{i}+\cdots+o_{k-2}\alpha_{i}^{k-2}+o_{k-1}\left(\alpha_{i}^{k-1}-\eta_{0}\alpha_{i}^{k}\right)\right),$$
it means that $$\boldsymbol{g}_{k-1}^{(0)}\in\mathrm{Span}_{\mathbb{F}_{q}}\left\{\boldsymbol{h}_{0,2}^{(0,2k)},\ldots,\boldsymbol{h}_{k-2,2}^{(0,2k)},\boldsymbol{h}_{k-1,2}^{(0,2k)}\right\},$$
then the code $\mathcal{C}_{k-1}\left(\boldsymbol{\alpha},\boldsymbol{v},\boldsymbol{\eta}\right)$ is self-orthogonal. 
\subsection{The case $1\leq \ell\leq n-k-1$} 
In this subsection, we discuss the existence for the self-orthogonal code $\mathcal{C}_{k-1}\left(\boldsymbol{\alpha},\boldsymbol{v},\boldsymbol{\eta}\right)$ when $1\leq \ell\leq n-k-1$. 

 Firstly, two sufficient and necessary conditions and a sufficient conditions for the code $\mathcal{C}_{k-1}\left(\boldsymbol{\alpha},\boldsymbol{v},\boldsymbol{\eta}\right)$ to be self-orthogonal are given.
\begin{theorem}\label{l=1n=2k+1yes1}
If $\ell=1$, $n=2k+1$, $1+\eta_{0}S_1+\eta_{1}S_{2}\neq 0$ and there exists some  $\lambda\in\mathbb{F}_{q}^{*}$ such that $\lambda u_{i}=v_{i}^{2}$ for $1 \leq i \leq n$, then the code $\mathcal{C}_{k-1}\left(\boldsymbol{\alpha},\boldsymbol{v},\boldsymbol{\eta}\right)$ is self-orthogonal if and only if $2\eta_{1}+\eta_{0}^{2}+2\eta_{0}\eta_{1}S_{1}+\eta_{1}^{2}S_{2}=0.$
\end{theorem}
{\bf Proof.} By Theorem \ref{+LPTGRSparitymatrix} (1), for $\ell=1$, $n=2k+1$ and $1+\sum\limits_{t=0}^{1}\eta_{t}S_{t+1}\neq 0$, the code $\mathcal{C}_{k-1}\left(\boldsymbol{\alpha},\boldsymbol{v},\boldsymbol{\eta}\right)$ has generator matrix $\boldsymbol{G}_{k,+}^{(1)}$ given in $(\ref{+LPTGRSgeneratormattrix})$ and the parity-check matrix  
$$
\boldsymbol{H}_{k+1,+,1}^{(1,2k+1)}=\begin{pmatrix}
	\frac{u_{1}}{v_{1}}&\cdots&\frac{u_{n}}{v_{n}}\\
	\frac{u_{1}}{v_{1}}\alpha_{1}&\cdots&\frac{u_{n}}{v_{n}}\alpha_{n}\\
	\vdots& &\vdots\\
	\frac{u_{1}}{v_{1}}\alpha_{1}^{k-2}&\cdots&\frac{u_{n}}{v_{n}}\alpha_{n}^{k-2}\\
	\frac{u_{1}}{v_{1}}\left(\alpha_{1}^{k-1}-\frac{\eta_{1}}{1+\sum\limits_{t=0}^{1}\eta_{t}S_{t+1}}\alpha_{1}^{k+1}\right)&\cdots&\frac{u_{n}}{v_{n}}\left(\alpha_{n}^{k-1}-\frac{\eta_{1}}{1+\sum\limits_{t=0}^{1}\eta_{t}S_{t+1}}\alpha_{n}^{k+1}\right)\\	\frac{u_{1}}{v_{1}}\left(\alpha_{1}^{k}-\frac{\eta_{0}+\eta_{1}S_{1}}{1+\sum\limits_{t=0}^{1}\eta_{t}S_{t+1}}\alpha_{1}^{k+1}\right)&\cdots&\frac{u_{n}}{v_{n}}\left(\alpha_{n}^{k}-\frac{\eta_{0}+\eta_{1}S_{1}}{1+\sum\limits_{t=0}^{1}\eta_{t}S_{t+1}}\alpha_{n}^{k+1}\right)
\end{pmatrix}.
$$
For convenience, we set $\boldsymbol{h}_{i,1}^{(1,2k+1)}$ denote the $(i+1)$-th row vector of $\boldsymbol{H}_{k+1,+,1}^{(1,2k+1)}$.

By definition, the code $\mathcal{C}_{k-1}\left(\boldsymbol{\alpha},\boldsymbol{v},\boldsymbol{\eta}\right)$ is self-orthogonal if and only if for any $0\leq i\leq k-1$,
$$\boldsymbol{g}_{i}^{(1)}\in\mathcal{C}_{k-1}^{\perp}\left(\boldsymbol{\alpha},\boldsymbol{v},\boldsymbol{\eta}\right).$$
It's easy to know that $\boldsymbol{g}_{i}(0\leq i\leq k-2)$ can be represented by $\boldsymbol{h}_{0,1}^{(1,2k+1)},\ldots,\boldsymbol{h}_{k-2,1}^{(1,2k+1)}$, i.e., $$\boldsymbol{g}_{i}\in\mathcal{C}_{k-1}^{\perp}\left(\boldsymbol{\alpha},\boldsymbol{v},\boldsymbol{\eta}\right)(0\leq i\leq k-2).$$ 
Then the code $\mathcal{C}_{k-1}\left(\boldsymbol{\alpha},\boldsymbol{v},\boldsymbol{\eta}\right)$ is self-orthogonal if and only if 
$$\boldsymbol{g}_{k-1}^{(1)}\in\mathcal{C}_{k-1}^{\perp}\left(\boldsymbol{\alpha},\boldsymbol{v},\boldsymbol{\eta}\right),$$
i.e., $$\boldsymbol{g}_{k-1}^{(1)}\in\mathrm{Span}_{\mathbb{F}_{q}}\left\{\boldsymbol{h}_{0,1}^{(1,2k+1)},\ldots,\boldsymbol{h}_{k-2,1}^{(1,2k+1)},\boldsymbol{h}_{k-1,1}^{(1,2k+1)},\boldsymbol{h}_{k,1}^{(1,2k+1)}\right\},$$
it means that for any $1\leq i\leq n$, there exists some $r_{j}(0\leq j\leq k)$ not all zero such that 
$$\begin{aligned}
	&v_{i}\left(\alpha_{i}^{k-1}+\eta_{0}\alpha_{i}^{k}+\eta_{1}\alpha_{i}^{k+1}\right)\\
	=&\frac{u_{i}}{v_{i}}\left(r_{0}+r_{1}\alpha_{i}+\cdots+r_{k-2}\alpha_{i}^{k-2}+r_{k-1}\left(\alpha_{i}^{k-1}-\frac{\eta_{1}}{1+\sum\limits_{t=0}^{1}\eta_{t}S_{t+1}}\alpha_{i}^{k+1}\right)\right.\\
	&+\left.r_{k}\left(\alpha_{i}^{k}-\frac{\eta_{0}+\eta_{1}S_{1}}{1+\sum\limits_{t=0}^{1}\eta_{t}S_{t+1}}\alpha_{i}^{k+1}\right)\right)
\end{aligned},$$
i.e.,
$$\begin{cases}
\frac{u_i}{v_i}r_{j}=0,&\ \text{if}\ 0\leq j\leq k-2;\\
\frac{u_i}{v_i}r_{k-1}=v_i;&\\
\frac{u_i}{v_i}r_{k}=v_i\eta_{0};&\\
-\frac{u_i}{v_i}\left(r_{k-1}\cdot\frac{\eta_{1}}{1+\sum\limits_{t=0}^{1}\eta_{t}S_{t+1}} +r_{k}\cdot \frac{\eta_{0}+\eta_{1}S_{1}}{1+\sum\limits_{t=0}^{1}\eta_{t}S_{t+1}}\right)=v_i\eta_{1},&\\
\end{cases}$$
namely, 
$$\begin{cases}
r_{j}=0,&\ \text{if}\ 0\leq j\leq k-2;\\
r_{k-1}=\frac{v_{i}^{2}}{u_i};&\\
r_{k}=\frac{v_{i}^{2}}{u_i}\cdot\eta_{0};&\\
\frac{-\eta_{1}}{1+\sum\limits_{t=0}^{1}\eta_{t}S_{t+1}} +\frac{-\eta_{0}\left(\eta_{0}+\eta_{1}S_{1}\right)}{1+\sum\limits_{t=0}^{1}\eta_{t}S_{t+1}}=\eta_{1},&\\
\end{cases}$$
it means that the code $\mathcal{C}_{k-1}\left(\boldsymbol{\alpha},\boldsymbol{v},\boldsymbol{\eta}\right)$ is self-orthogonal if and only if there exists some  $\lambda\in\mathbb{F}_{q}^{*}$ such that $\lambda u_{i}=v_{i}^{2}$ for $1 \leq i \leq n$ and 
$$\frac{-\eta_{1}}{1+\sum\limits_{t=0}^{1}\eta_{t}S_{t+1}} +\frac{-\eta_{0}\left(\eta_{0}+\eta_{1}S_{1}\right)}{1+\sum\limits_{t=0}^{1}\eta_{t}S_{t+1}}=\eta_{1},$$
i.e., $2\eta_{1}+\eta_{0}^{2}+2\eta_{0}\eta_{1}S_{1}+\eta_{1}^{2}S_{2}=0$.

From the above discussions, we complete the proof of Theorem $\ref{l=1n=2k+1yes1}$.

$\hfill\Box$ 

Similar to the proof of Theorem $\ref{l=1n=2k+1yes1}$,
one can obtain the following Theorem $\ref{l=1n=2k+1yes2}$.
\begin{theorem}\label{l=1n=2k+1yes2}
If $\ell=1$, $n=2k+1$, $1+\eta_{0}S_1+\eta_{1}S_{2}=0$ and there exists some  $\lambda\in\mathbb{F}_{q}^{*}$ such that $\lambda u_{i}=v_{i}^{2}$ for $1 \leq i \leq n$, then the code $\mathcal{C}_{k-1}\left(\boldsymbol{\alpha},\boldsymbol{v},\boldsymbol{\eta}\right)$ is self-orthogonal if and only if $\eta_{0}^2+\eta_{0}\eta_{1}S_{1}+\eta_{1}=0.$
\end{theorem}

Next,  we give two sufficient conditions for the code $\mathcal{C}_{k-1}\left(\boldsymbol{\alpha},\boldsymbol{v},\boldsymbol{\eta}\right)$ to be self-orthogonal.
\begin{theorem}\label{lgeq1n=2k+l+1yes1}
	Let $2\leq k=\frac{n-\ell-1}{2}$ with $\ell\geq 1$. If  the following two conditions hold simultaneously,
	
	$(1)$ there exists some  $\lambda\in\mathbb{F}_{q}^{*}$ such that $\lambda u_{i}=v_{i}^{2}$ for $1 \leq i \leq n$;
	
	$(2)$ $\sum\limits_{t=0}^{\ell}S_{t}\sum\limits_{i=t}^{\ell}\eta_{i}\eta_{\ell+t-i}=0,$
\\then the code $\mathcal{C}_{k-1}\left(\boldsymbol{\alpha},\boldsymbol{v},\boldsymbol{\eta}\right)$ is self-orthogonal. 
\end{theorem} 
{\bf Proof.} By directly computing and $\lambda u_{i}=v_{i}^{2}$ for $1 \leq i \leq n$, we have
\[
\boldsymbol{g}_{i}^{(\ell)}
\left(\boldsymbol{g}_{j}^{(\ell)}\right)^T
=
\begin{cases}
	\lambda\displaystyle\sum_{r=1}^{n}
	u_r\alpha_r^{i+j},
	& 0\le i,j\le k-2,\\
	\lambda\displaystyle\sum_{r=1}^{n}
	u_r\alpha_r^{i}
	\left(
	\alpha_r^{k-1}
	+\sum_{t=0}^{\ell}\eta_t\alpha_r^{k+t}
	\right),
	& 0\le i\le k-2,\ j=k-1,\\
	\lambda\displaystyle\sum_{r=1}^{n}
	u_r\alpha_r^{j}
	\left(
	\alpha_r^{k-1}
	+\sum_{t=0}^{\ell}\eta_t\alpha_r^{k+t}
	\right),
	& i=k-1,\ 0\le j\le k-2,\\
	\lambda\displaystyle\sum_{r=1}^{n}
	u_r
	\left(
	\alpha_r^{k-1}
	+\sum_{t=0}^{\ell}\eta_t\alpha_r^{k+t}
	\right)^2,
	& i=j=k-1.
\end{cases}
\]
By Lemma \ref{usapowersum} and $\sum\limits_{t=0}^{\ell}S_{t}\sum\limits_{i=t}^{\ell}\eta_{i}\eta_{\ell+t-i}=0,$, we know $\boldsymbol{g}_{i}^{(\ell)}
\left(\boldsymbol{g}_{j}^{(\ell)}\right)^T
=0$ for $0\leq i,j\leq k-1$, furthermore, $\boldsymbol{G}_{k,+}^{(\ell)}
\left(\boldsymbol{G}_{k,+}^{(\ell)}\right)^T=0,$ then the code $\mathcal{C}_{k-1}\left(\boldsymbol{\alpha},\boldsymbol{v},\boldsymbol{\eta}\right)$ is self-orthogonal. 

$\hfill\Box$ 

By Definition \ref{completesymmetricpolynomial} and Lemma \ref{mxroot}, it's easy to know that if $\alpha_{1},\ldots,\alpha_{n}$ are $n$ distinct roots of $x^n-\mu\in\mathbb{F}_{q}[x],$ where  $n\mid (q-1)$ and $\mu\in\mathbb{F}_{q}^{*}$ with $\mathrm{ord}(\mu)\mid \frac{q-1}{n}$ , then $S_{N}=0$ for any $1\leq N\leq \ell<n$, furthermore, we immediately get the following corollary. 
\begin{corollary}\label{lgeq1n=2k+l+1yes2}
Let $2\leq k=\frac{n-\ell-1}{2}$ with $\ell\geq 1$  and $n\mid (q-1)$, $\alpha_{1},\ldots,\alpha_{n}$ are $n$ distinct roots of $x^n-\mu\in\mathbb{F}_{q}[x],$ where $\mu\in\mathbb{F}_{q}^{*}$ with $\mathrm{ord}(\mu)\mid \frac{q-1}{n}$. If the following two conditions hold simultaneously,

$(1)$ there exists some  $\lambda\in\mathbb{F}_{q}^{*}$ such that $\lambda u_{i}=v_{i}^{2}$ for $1 \leq i \leq n$;

$(2)$ $\sum\limits_{i=0}^{\ell}\eta_{i}\eta_{\ell-i}=0,$\\
then the code $\mathcal{C}_{k-1}\left(\boldsymbol{\alpha},\boldsymbol{v},\boldsymbol{\eta}\right)$ is self-orthogonal. 
\end{corollary}
\begin{remark}
For Corollary \ref{lgeq1n=2k+l+1yes2}, if $\ell=1$, then $\sum\limits_{i=0}^{\ell}\eta_{i}\eta_{\ell-i}=0$ if and only if $2\eta_{0}\eta_{1}=0$, i.e, $\mathrm{Char}\left(\mathbb{F}_{q}\right)=2$ or $\eta_{0}=0$.
\end{remark} 

\begin{theorem}\label{lgeq1ngeq2k+2l+2yes}
If $2\leq k\leq\frac{n-2\ell-2}{2}$ and there exists some  $\lambda\in\mathbb{F}_{q}^{*}$ such that $\lambda u_{i}=v_{i}^{2}$ for $1 \leq i \leq n$, then the code $\mathcal{C}_{k-1}\left(\boldsymbol{\alpha},\boldsymbol{v},\boldsymbol{\eta}\right)$ is self-orthogonal.
\end{theorem}
{\bf Proof}. By Theorem \ref{+LPTGRSparitymatrix}, we know that the code $\mathcal{C}_{k-1}\left(\boldsymbol{\alpha},\boldsymbol{v},\boldsymbol{\eta}\right)$ has the generator matrix $\boldsymbol{G}_{k,+}^{(\ell)}$ given in $(\ref{+LPTGRSgeneratormattrix})$ and the parity-check matrix $\boldsymbol{H}_{n-k,+,1}^{(\ell,n)}$ or $\boldsymbol{H}_{n-k,+,2}^{(\ell,n)}$ given in $(\ref{paritycheckmatrix1})$ or $(\ref{paritymatrix2})$, respectively. By $k\leq\frac{n-2\ell-2}{2}$, we have $k+\ell\leq n-k-(\ell+2)$, and then $\boldsymbol{g}_{i}^{(\ell)}(0\leq i\leq k-1)$ can be represented by $\boldsymbol{h}_{0,1}^{(\ell,n)},\ldots,\boldsymbol{h}_{n-k-(\ell+2),1}^{(\ell,n)}$ or $\boldsymbol{h}_{0,2}^{(\ell,n)},\ldots,\boldsymbol{h}_{n-k-(\ell+2),2}^{(\ell,n)}$, respectively, thus 
$$\boldsymbol{g}_{i}^{(\ell)}\in\mathrm{Span}\left\{\boldsymbol{h}_{0,1}^{(\ell,n)},\ldots,\boldsymbol{h}_{n-k-(\ell+2),1}^{(\ell,n)},\boldsymbol{h}_{n-k-(\ell+1),1}^{(\ell,n)},\ldots,\boldsymbol{h}_{n-k-1,1}^{(\ell,n)}\right\}$$
or 
$$\boldsymbol{g}_{i}^{(\ell)}\in\mathrm{Span}\left\{\boldsymbol{h}_{0,2}^{(\ell,n)},\ldots,\boldsymbol{h}_{n-k-(\ell+2),2}^{(\ell,n)},\boldsymbol{h}_{n-k-\ell,2}^{(\ell,n)},\ldots,\boldsymbol{h}_{n-k,2}^{(\ell,n)}\right\},$$
i.e., the code $\mathcal{C}_{k-1}\left(\boldsymbol{\alpha},\boldsymbol{v},\boldsymbol{\eta}\right)$ is self-orthogonal.
 
In the  same proof  as that of Theorem $\ref{lgeq1ngeq2k+2l+2yes}$, it's easy to get the following 
\begin{theorem}\label{l=0ngeq2k+2}
For the code $\mathcal{C}_{k-1}\left(\boldsymbol{\alpha},\boldsymbol{v},\boldsymbol{\eta}\right)$ with $\ell=0$, if $n\geq2k+2\geq 6$ and there exists some  $\lambda\in\mathbb{F}_{q}^{*}$ such that $\lambda u_{i}=v_{i}^{2}$ for $1 \leq i \leq n$, then the code $\mathcal{C}_{k-1}\left(\boldsymbol{\alpha},\boldsymbol{v},\boldsymbol{\eta}\right)$ is self-orthogonal.
\end{theorem}
\begin{remark}
By taking $\lambda=1$ in Theorem $\ref{l=0ngeq2k+2}$, the corresponding result is just the case $h=k-1$ of Theorem 6 in \cite{LiuTGRSDCC2021}. 
\end{remark} 

Finally,  three sufficient conditions for the code $\mathcal{C}_{k-1}\left(\boldsymbol{\alpha},\boldsymbol{v},\boldsymbol{\eta}\right)$ not to be self-orthogonal are given.
\begin{theorem}\label{notl=1n=2k+1yes2}
	If there exists some  $\lambda\in\mathbb{F}_{q}^{*}$ such that $\lambda u_{i}=v_{i}^{2}$ for $1 \leq i \leq n$ and one of the following conditions is satisfied, then the code $\mathcal{C}_{k-1}\left(\boldsymbol{\alpha},\boldsymbol{v},\boldsymbol{\eta}\right)$ is not self-orthogonal.
	
	$(1)$ $n=2k$ and $\ell\geq 1$;
	
	$(2)$ $n=2k+1$ and $\ell\geq 2$;
	
	$(3)$ $n=2k+3$ and $\ell=1$.
\end{theorem}
{\bf Proof.} By Definition \ref{LPTGRSdefinition}, we know that the code $\mathcal{C}_{k-1}\left(\boldsymbol{\alpha},\boldsymbol{v},\boldsymbol{\eta}\right)$ has the generator matrix $\boldsymbol{G}_{k,+}^{(\ell)}$ given in $(\ref{+LPTGRSgeneratormattrix})$. 

{\bf (1)} For $1+\sum\limits_{t=0}^{\ell}\eta_{t}S_{t+1}\neq 0$, by $n=2k$, by Theorem \ref{+LPTGRSparitymatrix} (1), we know that the code $\mathcal{C}_{k-1}\left(\boldsymbol{\alpha},\boldsymbol{v},\boldsymbol{\eta}\right)$ has the following parity-matrix
\begin{equation} 
	\boldsymbol{H}_{k,+,1}^{(\ell,2k)}=\begin{pmatrix}
		\cdots&\frac{u_{j}}{v_{j}}&\cdots\\
		\cdots&\frac{u_{j}}{v_{j}}\alpha_{j}&\cdots\\
		\vdots&\vdots&\vdots\\
		\cdots&\frac{u_{j}}{v_{j}}\alpha_{j}^{k-\ell-2}&\cdots\\
		\cdots&\frac{u_{j}}{v_{j}}\left(\alpha_{j}^{k-\ell-1}-\varTheta_{k-\ell-1}\alpha_{j}^{k}\right)&\cdots\\ 
		\vdots&\vdots&\vdots\\
		\cdots&\frac{u_{j}}{v_{j}}\left(\alpha_{j}^{k-1}-\varTheta_{k-1}\alpha_{j}^{k}\right)&\cdots
	\end{pmatrix}_{k\times 2k}=\begin{pmatrix}
		\boldsymbol{h}_{0,1}^{(\ell,2k)}\\
		\boldsymbol{h}_{1,1}^{(\ell,2k)}\\
		\vdots\\
		\boldsymbol{h}_{k-\ell-2,1}^{(\ell,2k)}\\
		\boldsymbol{h}_{k-\ell-1,1}^{(\ell,2k)}\\
		\vdots\\
		\boldsymbol{h}_{k-1,1}^{(\ell,2k)}
	\end{pmatrix}.
\end{equation} Note that $\ell\geq 1$, we have $k+\ell\geq k+1>k$. If $\mathcal{C}_{k-1}\left(\boldsymbol{\alpha},\boldsymbol{v},\boldsymbol{\eta}\right)$ is self-orthogonal, then there exist $e_i(0\leq i\leq k-1)$ such that  $$\boldsymbol{g}_{k-1}^{(\ell)}=v_i\left(\boldsymbol{\alpha}_{i}^{k-1}+\sum\limits_{t=0}^{\ell}\eta_t\boldsymbol{\alpha}_{i}^{k+t}\right)=e_{0}\boldsymbol{h}_{0,1}^{(\ell,2k)}+e_{1}\boldsymbol{h}_{1,1}^{(\ell,2k)}+\cdots+e_{k-1}\boldsymbol{h}_{k-1,1}^{(\ell,2k)},$$
i.e., the polynomial
$$\begin{aligned}
&\lambda\left(x^{k-1}+\sum\limits_{i=0}^{\ell}\eta_ix^{k+t}\right)\\
-&\left(e_{0}+e_{1}x+\cdots+e_{k-\ell-2}x^{k-\ell-2}+e_{k-\ell-1}\left(x^{k-\ell-1}-\Theta_{k-\ell-1}x^{k}\right)+\cdots+e_{k-1}\left(x^{k-1}-\Theta_{k-1}x^{k}\right)\right),  
\end{aligned}$$
has $n$ distinct roots, which is contradict with $n=2k>k+\ell$, since $0\leq \ell\leq n-k-1$, i.e., $$\begin{aligned}
	&\lambda\left(x^{k-1}+\sum\limits_{i=0}^{\ell}\eta_ix^{k+t}\right)\\
	=&\left(e_{0}+e_{1}x+\cdots+e_{k-\ell-2}x^{k-\ell-2}+e_{k-\ell-1}\left(x^{k-\ell-1}-\Theta_{k-\ell-1}x^{k}\right)+\cdots+e_{k-1}\left(x^{k-1}-\Theta_{k-1}x^{k}\right)\right),  
\end{aligned}$$ 
By comparing the coefficient of \(x^{k+\ell}\) yields
\(\lambda=0\), a contradiction. Therefore, $$\boldsymbol{g}_{k-1}^{(\ell)}\notin\mathrm{Span}_{\mathbb{F}_{q}}\left\{\boldsymbol{h}_{0,1}^{(\ell,2k)},\ldots,\boldsymbol{h}_{k-1,1}^{(\ell,2k)}\right\},
$$ i.e., the code $\mathcal{C}_{k-1}\left(\boldsymbol{\alpha},\boldsymbol{v},\boldsymbol{\eta}\right)$ is not self-orthogonal. 

For $1+\sum\limits_{t=0}^{\ell}\eta_{t}S_{t+1}=0$, in the  same proof  as that of the above, we can complete the corresponding proof.

{\bf (2)} For $1+\sum\limits_{t=0}^{\ell}\eta_{t}S_{t+1}\neq 0$, by $n=2k+1$, we know that the code $\mathcal{C}_{k-1}\left(\boldsymbol{\alpha},\boldsymbol{v},\boldsymbol{\eta}\right)$ has the following parity-matrix
$$\boldsymbol{H}_{k+1,+,1}^{(\ell,2k+1)}=\begin{pmatrix}
	\cdots&\frac{u_{j}}{v_{j}}&\cdots\\
	\cdots&\frac{u_{j}}{v_{j}}\alpha_{j}&\cdots\\
	\vdots&\vdots&\vdots\\
	\cdots&\frac{u_{j}}{v_{j}}\alpha_{j}^{k-\ell-1}&\cdots\\
	\cdots&\frac{u_{j}}{v_{j}}\left(\alpha_{j}^{k-\ell}-\varTheta_{k-\ell}\alpha_{j}^{k+1}\right)&\cdots\\ 
	\vdots&\vdots&\vdots\\
	\cdots&\frac{u_{j}}{v_{j}}\left(\alpha_{j}^{k}-\varTheta_{k}\alpha_{j}^{k+1}\right)&\cdots
\end{pmatrix}_{(k+1)\times (2k+1)}=\begin{pmatrix}
	\boldsymbol{h}_{0,1}^{(\ell,2k+1)}\\
	\boldsymbol{h}_{1,1}^{(\ell,2k+1)}\\
	\vdots\\
	\boldsymbol{h}_{k-\ell-1,1}^{(\ell,2k+1)}\\
	\boldsymbol{h}_{k-\ell,1}^{(\ell,2k+1)}\\
	\vdots\\
	\boldsymbol{h}_{k,1}^{(\ell,2k+1)}
\end{pmatrix}.
$$
Note that $\ell\geq 2$, we have $k+\ell\geq k+2>k+1$, in the  same proof  as that of Theorem \ref{notl=1n=2k+1yes2} (1), one can prove that the code $\mathcal{C}_{k-1}\left(\boldsymbol{\alpha},\boldsymbol{v},\boldsymbol{\eta}\right)$ is not self-orthogonal. 

For $1+\sum\limits_{t=0}^{\ell}\eta_{t}S_{t+1}=0$, in the  same proof  as that of the above, we can complete the corresponding proof.

{\bf (3)} In the  same idea as that of Theorem \ref{notl=1n=2k+1yes2} (1)-(2), one can get the corresponding contradiction. Then the code $\mathcal{C}_{k-1}\left(\boldsymbol{\alpha},\boldsymbol{v},\boldsymbol{\eta}\right)$ is not self-orthogonal.

From the above discussions, we complete the proof of Theorem \ref{notl=1n=2k+1yes2}. 

$\hfill\Box$ 
\subsection{Applications to the LCD MDS code $\mathcal{C}\left(\alpha,v,\eta\right)$}
In 2025, for the $(\mathcal{L},\mathcal{P})$-TGRS code with the most general
form coefficient matrix $\boldsymbol{B}$, Hu et al. gave a necessary and sufficient condition  for this code to be MDS and gave the corresponding examples. Since the code $\mathcal{C}_{k-1}\left(\boldsymbol{\alpha},\boldsymbol{v},\boldsymbol{\eta}\right)$ is a special type of $(\mathcal{L},\mathcal{P})$-TGRS codes, then we can assume that the code $\mathcal{C}_{k-1}\left(\boldsymbol{\alpha},\boldsymbol{v},\boldsymbol{\eta}\right)$ is MDS.

In this subsection, we give some constructions of LCD MDS codes basing on the self-orthogonal code $\mathcal{C}_{k-1}\left(\boldsymbol{\alpha},\boldsymbol{v},\boldsymbol{\eta}\right)$ given in Theorems \ref{l=0yes}-\ref{lgeq1ngeq2k+2l+2yes}. 

By Lemma \ref{selforthogonalMDSLCD} and Theorems \ref{l=0yes}-\ref{lgeq1ngeq2k+2l+2yes}, it's easy to obtain the following 
\begin{theorem}\label{LCDMDS}
	For the MDS code $\mathcal{C}_{k-1}\left(\boldsymbol{\alpha},\boldsymbol{v},\boldsymbol{\eta}\right)$, if there exists some  $\lambda\in\mathbb{F}_{q}^{*}$ such that $\lambda u_{i}=v_{i}^{2}$ for $1 \leq i \leq n$ and one of the following conditions holds, then for any $\beta\in\mathbb{F}_{q}\backslash\left\{0,-1,1\right\}$, the code $\mathcal{C}\left(\boldsymbol{\alpha},\boldsymbol{v}^{\prime},\boldsymbol{\eta}\right)$ with $\boldsymbol{v}^{\prime}=\left(v_1,\ldots,v_k,\beta v_{k+1},\ldots,\beta v_{n}\right)$ is an LCD MDS code. 
	\begin{enumerate}[label={$(\theenumi)$}] 
		\item $\ell=0$, $n=2k$ with $k\geq 2$, $\sum\limits_{i=1}^{n}\alpha_{i}=0$ and $\mathrm{Char}\left(\mathbb{F}_{q}\right)=2$;
		\item $\ell=1$, $n=2k+1$, $1+\eta_{0}S_1+\eta_{1}S_{2}\neq 0$, and $2\eta_{1}+\eta_{0}^{2}+2\eta_{0}\eta_{1}S_{1}+\eta_{1}^{2}S_{2}=0$;
		\item $\ell=1$, $n=2k+1$, $1+\eta_{0}S_1+\eta_{1}S_{2}=0$, and $\eta_{0}^2+\eta_{0}\eta_{1}S_{1}+\eta_{1}=0;$
		\item $\ell\geq 1$, $n=2k+\ell+1$, and $\sum\limits_{t=0}^{\ell}S_{t}\sum\limits_{i=t}^{\ell}\eta_{i}\eta_{\ell+t-i}=0;$
		\item $\ell\geq 0,$ $n=2k+2\ell+2.$
	\end{enumerate}
\end{theorem}
\section{The NMDS or non-GRS code $\mathcal{C}\left(\alpha,v,\eta\right)$}\label{sec5}
In this section, for the code $\mathcal{C}_{k-1}\left(\boldsymbol{\alpha},\boldsymbol{v},\boldsymbol{\eta}\right)$, we give a necessary and sufficient condition for the code $\mathcal{C}_{k-1}\left(\boldsymbol{\alpha},\boldsymbol{v},\boldsymbol{\eta}\right)$ to be NMDS, and prove that the code  $\mathcal{C}_{k-1}\left(\boldsymbol{\alpha},\boldsymbol{v},\boldsymbol{\eta}\right)$ is non-GRS when $2k>n$. 
\subsection{The NMDS code $\mathcal{C}\left(\alpha,v,\eta\right)$}
In this subsection, we give a necessary and sufficient condition for the code $\mathcal{C}_{k-1}\left(\boldsymbol{\alpha},\boldsymbol{v},\boldsymbol{\eta}\right)$ to be NMDS.
\begin{theorem}\label{+LPTGRSNMDScondition}
For any $k$-subset $\mathcal{I}\subseteq\left\{1,\ldots,n\right\}$, let $\prod\limits_{j\in\mathcal{I}}\left(x-\alpha_{j}\right)=\sum\limits_{j\in\mathcal{I}}c_jx^{k-j}$, $c_j=0$ for $j>k$, $\boldsymbol{\beta}_{t}=\left(c_{t+1},\ldots,c_{2},c_{1}\right)$, $\boldsymbol{\gamma}_{t}^{T}=\left(1,0,\ldots,0\right)\in\mathbb{F}_{q}^{t+1}$ and 
	$$\boldsymbol{A}_{\mathcal{I},t}=\begin{pmatrix}
		1& & & \\
		c_1&1& & \\
		c_2&c_1&1& & \\
		\vdots&\vdots&\ddots&\ddots\\
		c_t&c_{t-1}&\cdots&c_{1}&1\\
	\end{pmatrix}.$$
	Then the code $\mathcal{C}_{k-1}\left(\boldsymbol{\alpha},\boldsymbol{v},\boldsymbol{\eta}\right)$ is NMDS if and only if the following two conditions hold simultaneously,
	
	$(1)$ $\boldsymbol{\eta}\notin \Omega=\left\{\boldsymbol{\eta}\in\mathbb{F}_{q}^{\ell+1}\backslash\left\{\boldsymbol{0}\right\}\Bigm|\text{any}\  k\text{-subset}\ \mathcal{I}\subseteq\left\{1,\ldots,n\right\},  1-\sum\limits_{t=0}^{\ell}\eta_{t}\boldsymbol{\beta}_{t}\boldsymbol{A}_{\mathcal{I},t}^{-1}\boldsymbol{\gamma}_{t}^{T}\neq 0\right\};$
	
	$(2)$ for any $(k+1)$-subset $\mathcal{J}\subseteq\left\{1,\ldots,n\right\}$, there exists some $k$-subset $\mathcal{I}\subseteq\mathcal{J}$ such that 
	$$1-\sum\limits_{t=0}^{\ell}\eta_{t}\boldsymbol{\beta}_{t}\boldsymbol{A}_{\mathcal{I},t}^{-1}\boldsymbol{\gamma}_{t}^{T}\neq 0.$$
\end{theorem}
{\bf Proof}. Note that $\boldsymbol{G}_{k,+}^{(\ell)}$ given in (\ref{+LPTGRSgeneratormattrix}) is the generator matrix of the code $\mathcal{C}_{k-1}\left(\boldsymbol{\alpha},\boldsymbol{v},\boldsymbol{\eta}\right)$, then by Lemma \ref{NMDScondition}, the code $\mathcal{C}_{k-1}\left(\boldsymbol{\alpha},\boldsymbol{v},\boldsymbol{\eta}\right)$ is NMDS if and only if the following conditions hold simultaneously,

$(i)$ any $k-1$ columns of  $\boldsymbol{G}_{k,+}^{(\ell)}$ are $\mathbb{F}_q$-linearly independent;

$(ii)$ there exist $k$ columns of $\boldsymbol{G}_{k,+}^{(\ell)}$ that are $\mathbb{F}_q$-linearly dependent;

$(iii)$ For any $k+1$ columns of $\boldsymbol{G}_{k,+}^{(\ell)}$, there exist $k$ columns $\mathbb{F}_q$-linearly independent.

{\bf For $(i)$,} without loss of generality, the submatrix $\boldsymbol{K}$ consisted of any $k-1$ columns in  $\boldsymbol{G}_{k,+}^{(\ell)}$ has the following form 
$$\small\boldsymbol{K}=\begin{pmatrix}
	v_1&\cdots&v_{k-1}\\
	v_1\alpha_{1}&\cdots&v_{k-1}\alpha_{k-1}\\
	\vdots& &\vdots\\
	v_1\alpha_{1}^{k-2}&\cdots&v_{k-1}\alpha_{k-1}^{k-2}\\
	v_1\left(\alpha_{1}^{k-1}+\sum\limits_{t=0}^{\ell}\eta_{t}\alpha_{1}^{k+t}\right)&\cdots&v_{k-1}\left(\alpha_{k-1}^{k-1}+\sum\limits_{t=0}^{\ell}\eta_{t}\alpha_{k-1}^{k+t}\right)
\end{pmatrix}_{k\times (k-1)}.$$
It's easy to know that $\rank(\boldsymbol{K})\leq k-1 $. Note that $\begin{pmatrix}
	v_{1}&\cdots&v_{k-1}\\
	v_{1}\alpha_{1}&\cdots&v_{k-1}\alpha_{k-1}\\
	\vdots& &\vdots\\
	v_{1}\alpha_{1}^{k-2}&\cdots&v_{k-1}\alpha_{k-1}^{k-2}
\end{pmatrix}$ is a $(k-1)\times (k-1)$ non-zero minor of $\boldsymbol{K}$, then $\rank(\boldsymbol{K})=k-1,$ i.e., any $k-1$ columns of  $\boldsymbol{G}_{k,+}^{(\ell)}$ are $\mathbb{F}_q$-linearly independent.

\textbf{For $(ii)$,} without loss of generality, it's easy to know that any $k\times k$ minor of  $\boldsymbol{G}_{k,+}^{(\ell)}$ has the following form
$$\small\begin{aligned}
	&\det\begin{pmatrix}
		v_{1}&\cdots&v_{k}\\
		v_{1}\alpha_{1}&\cdots&v_{k}\alpha_{k}\\
		\vdots& &\vdots\\
		v_{1}\alpha_{1}^{k-2}&\cdots&v_{k}\alpha_{k}^{k-2}\\
		v_{1}\left(\alpha_{1}^{k-1}+\sum\limits_{t=0}^{\ell}\eta_{t}\alpha_{1}^{k+t}\right)&\cdots&v_{k}\left(\alpha_{k}^{k-1}+\sum\limits_{t=0}^{\ell}\eta_{t}\alpha_{k}^{k+t}\right)
	\end{pmatrix}\\
	=&\prod\limits_{i=1}^{k}v_{i}\cdot\det\begin{pmatrix}
		1&\cdots&1\\
		\alpha_{1}&\cdots&\alpha_{k}\\
		\vdots& &\vdots\\
		\alpha_{1}^{k-2}&\cdots&\alpha_{k}^{k-2}\\
		\alpha_{1}^{k-1}&\cdots&\alpha_{k}^{k-1}
	\end{pmatrix}+\prod\limits_{i=1}^{k}v_{i}\cdot\sum\limits_{t=0}^{\ell}\eta_{t}\cdot\det\begin{pmatrix}
		1&\cdots&1\\
		\alpha_{1}&\cdots&\alpha_{k}\\
		\vdots& &\vdots\\
		\alpha_{1}^{k-2}&\cdots&\alpha_{k}^{k-2}\\
		\alpha_{1}^{k+t}&\cdots&\alpha_{k}^{k+t}
	\end{pmatrix}\\
	=&\prod\limits_{i=1}^{k}v_{i}\cdot\left(1-\sum\limits_{t=0}^{\ell}\eta_{t}\boldsymbol{\beta}_{t}\boldsymbol{A}_{\mathcal{I},t}^{-1}\boldsymbol{\gamma}_{t}^{T}\right)\prod\limits_{1\leq j<i\leq k}\left(\alpha_{i}-\alpha_{j}\right),
\end{aligned}$$
and so any $k$ columns of  $\boldsymbol{G}_{k,+}^{(\ell)}$ are $\mathbb{F}_q$-linearly independent if and only if 
$$\boldsymbol{\eta}\in \Omega=\left\{\boldsymbol{\eta}\in\mathbb{F}_{q}^{\ell+1}\backslash\left\{\boldsymbol{0}\right\}\Bigm|\forall\  k\text{-subset}\ \mathcal{I}\subseteq\left\{1,\ldots,n\right\},  1-\sum\limits_{t=0}^{\ell}\eta_{t}\boldsymbol{\beta}_{t}\boldsymbol{A}_{\mathcal{I},t}^{-1}\boldsymbol{\gamma}_{t}^{T}\neq 0\right\}.$$
Furthermore, there exist $k$ columns of $\boldsymbol{G}_{k,+}^{(\ell)}$ $\mathbb{F}_q$-linearly dependent if and only if 
$$\boldsymbol{\eta}\notin \Omega=\left\{\boldsymbol{\eta}\in\mathbb{F}_{q}^{\ell+1}\backslash\left\{\boldsymbol{0}\right\}\Bigm|\forall\  k\text{-subset}\ \mathcal{I}\subseteq\left\{1,\ldots,n\right\},  1-\sum\limits_{t=0}^{\ell}\eta_{t}\boldsymbol{\beta}_{t}\boldsymbol{A}_{\mathcal{I},t}^{-1}\boldsymbol{\gamma}_{t}^{T}\neq 0\right\}.$$

\textbf{For $(iii)$,} without loss of generality, the submatrix $\boldsymbol{L}$ consisting of any $k+1$ columns in  $\boldsymbol{G}_{k,+}^{(\ell)}$ has the following form 
$$\boldsymbol{L}=\begin{pmatrix}
	v_{1}&\cdots&v_{k+1}\\
	v_{1}\alpha_{1}&\cdots&v_{k+1}\alpha_{k+1}\\
	\vdots& &\vdots\\
	v_{1}\alpha_{1}^{k-2}&\cdots&v_{k+1}\alpha_{k+1}^{k-2}\\
	v_{1}\left(\alpha_{1}^{k-1}+\sum\limits_{t=0}^{\ell}\eta_{t}\alpha_{1}^{k+t}\right)&\cdots&v_{k+1}\left(\alpha_{k+1}^{k-1}+\sum\limits_{t=0}^{\ell}\eta_{t}\alpha_{k+1}^{k+t}\right)
\end{pmatrix}_{k\times (k+1)}.$$
It's easy to know $\rank(\boldsymbol{L})\leq k$. Then $\rank(\boldsymbol{L})=k$ if and only if there exists some $k\times k$ non-zero minor in $L$, i.e., for any $(k+1)$-subset $\mathcal{J}\subseteq\left\{1,\ldots,n\right\}$, there exists some $k$-subset $\mathcal{I}\subseteq\mathcal{J}$ such that 
$$1-\sum\limits_{t=0}^{\ell}\eta_{t}\boldsymbol{\beta}_{t}\boldsymbol{A}_{\mathcal{I},t}^{-1}\boldsymbol{\gamma}_{t}^{T}\neq 0.$$

From the above discussions, we complete the proof of Theorem $\ref{+LPTGRSNMDScondition}$. 

$\hfill\Box$ 

\subsection{The non-GRS code $\mathcal{C}_{k-1}\left(\alpha,v,\eta\right)$}
 In this subsection, by calculating the dimension of the Schur square for the dual code $\mathcal{C}_{k-1}^{\perp}\left(\boldsymbol{\alpha},\boldsymbol{1},\boldsymbol{\eta}\right)$, we show 
 that the code $\mathcal{C}_{k-1}\left(\boldsymbol{\alpha},\boldsymbol{1},\boldsymbol{\eta}\right)$ is non-RS for some cases. 
\begin{theorem}\label{nonRS2kn}
For $2k>n>k+\ell+2$ and $1+\sum\limits_{t=0}^{\ell}\eta_{t}S_{t+1}\neq0$, or $2k>n\geq k+\ell+2\geq k+3$ and $1+\sum\limits_{t=0}^{\ell}\eta_{t}S_{t+1}=0$, the code $\mathcal{C}_{k-1}\left(\boldsymbol{\alpha},\boldsymbol{1},\boldsymbol{\eta}\right)$ is non-GRS. 
\end{theorem}
{\bf Proof}. For convenience, we denote $\boldsymbol{u}=\left(u_{1},\ldots,u_{n}\right)$ and $\boldsymbol{\alpha}^{z}=\left(\alpha_{1}^{z},\ldots,\alpha_{n}^{z}\right)$ for any non-negative integer $z$. Then by Theorem \ref{+LPTGRSparitymatrix}, it's easy to get 
$$\begin{aligned}
\mathcal{C}_{k-1}^{\perp}\left(\boldsymbol{\alpha},\boldsymbol{1},\boldsymbol{\eta}\right)=&\begin{cases}
\left\langle\boldsymbol{u}\star\boldsymbol{\alpha}^{i},\boldsymbol{u}\star\left(\boldsymbol{\alpha}^{s}-\varTheta_{s}\boldsymbol{\alpha}^{n-k}\right)\right\rangle, &\text{if}\  1+\sum\limits_{t=0}^{\ell}\eta_{t}S_{t+1}\neq 0;\\
\left\langle\boldsymbol{u}\star\boldsymbol{\alpha}^{i},\boldsymbol{u}\star\left(\boldsymbol{\alpha}^{t}-\Omega_{t}\boldsymbol{\alpha}^{n-k-(\ell+1) }\right),\boldsymbol{u}\star\boldsymbol{\alpha}^{n-k}\right\rangle, &\text{if}\  1+\sum\limits_{t=0}^{\ell}\eta_{t}S_{t+1}=0\ \text{and}\ \ell\geq 1,\\
\end{cases}
\end{aligned}$$
where $0\leq i\leq n-k-\left(\ell+2\right),n-k-\left(\ell+1\right)\leq s\leq n-k-1$ and $n-k-\ell\leq t\leq n-k-1$.

Firstly, for $1+\sum\limits_{t=0}^{\ell}\eta_{t}S_{t+1}\neq 0$,  by Definition \ref{schurproduct}, we have
$$
\begin{aligned}
&\left(\mathcal{C}_{k-1}^{\perp}\left(\boldsymbol{\alpha},\boldsymbol{1},\boldsymbol{\eta}\right)\right)^2\\
=&\left\langle\boldsymbol{u}\star\boldsymbol{\alpha}^{i},\boldsymbol{u}\star\left(\boldsymbol{\alpha}^{s_1}-\varTheta_{s_1}\boldsymbol{\alpha}^{n-k}\right)\right\rangle\star\left\langle\boldsymbol{u}\star\boldsymbol{\alpha}^{j},\boldsymbol{u}\star\left(\boldsymbol{\alpha}^{s_2}-\varTheta_{s_2}\boldsymbol{\alpha}^{n-k}\right)\right\rangle\\
&\left(i,j\in\left\{0,1,\ldots,n-k-\left(\ell+2\right)\right\}, s_1,s_2\in\left\{ n-k-\left(\ell+1\right),\ldots, n-k-1\right\}\right)\\
=&\left\langle\boldsymbol{u}^{2}\star\boldsymbol{\alpha}^{i+j},\boldsymbol{u}^{2}\star\boldsymbol{\alpha}^{i}\star\left(\boldsymbol{\alpha}^{s_2}-\varTheta_{s_2}\boldsymbol{\alpha}^{n-k}\right),\boldsymbol{u}^{2}\star\left(\boldsymbol{\alpha}^{s_1}-\varTheta_{s_1}\boldsymbol{\alpha}^{n-k}\right)\star\boldsymbol{\alpha}^{j},\right.\\
&\left.
\boldsymbol{u}^{2}\star\left(\boldsymbol{\alpha}^{s_1}-\varTheta_{s_1}\boldsymbol{\alpha}^{n-k}\right)\star\left(\boldsymbol{\alpha}^{s_2}-\varTheta_{s_2}\boldsymbol{\alpha}^{n-k}\right)\right\rangle\\
&\left(i,j\in\left\{0,1,\ldots,n-k-\left(\ell+2\right)\right\}, s_1,s_2\in\left\{ n-k-\left(\ell+1\right),\ldots, n-k-1\right\}\right)\\
=&\left\langle\boldsymbol{u}^{2}\star\boldsymbol{\alpha}^{i+j},\boldsymbol{u}^{2}\star\left(\boldsymbol{\alpha}^{i+s_2}-\varTheta_{s_2}\boldsymbol{\alpha}^{n-k+i}\right),\boldsymbol{u}^{2}\star\left(\boldsymbol{\alpha}^{s_1+j}-\varTheta_{s_1}\boldsymbol{\alpha}^{n-k+j}\right),\right.\\
&\left.\boldsymbol{u}^{2}\star\left(\boldsymbol{\alpha}^{s_1+s_2}-\varTheta_{s_1}\boldsymbol{\alpha}^{n-k+s_2}-\varTheta_{s_2}\boldsymbol{\alpha}^{n-k+s_1}+\varTheta_{s_1}\varTheta_{s_2}\boldsymbol{\alpha}^{2n-2k}\right)\right\rangle\\
&\left(i,j\in\left\{0,1,\ldots,n-k-\left(\ell+2\right)\right\}, s_1,s_2\in\left\{ n-k-\left(\ell+1\right),\ldots, n-k-1\right\}\right).\\
\end{aligned}
$$
By $2k>n>k+\ell+2$, we have $$n-k-\ell-2< 2n-2k-2\ell-4$$ and $$2n-2k-1<n-1,$$ then
$$\boldsymbol{u}^{2}\star\boldsymbol{\alpha}^{\color{red}{0}},\boldsymbol{u}^{2}\star\boldsymbol{\alpha}^{\color{red}{1}},\ldots\boldsymbol{u}^{2}\star\boldsymbol{\alpha}^{\color{red}{n-k-\ell-2}}, \boldsymbol{u}^{2}\star\left(\boldsymbol{\alpha}^{\color{red}{n-k-\ell-1}}-\varTheta_{n-k-\ell-1}\boldsymbol{\alpha}^{n-k}\right),\ldots,\boldsymbol{u}^{2}\star\left(\boldsymbol{\alpha}^{\color{red}{n-k-1}}-\varTheta_{n-k-1}\boldsymbol{\alpha}^{n-k}\right),$$
$$\boldsymbol{u}^{2}\star\left(\boldsymbol{\alpha}^{\color{red}{n-k}}-\varTheta_{n-k-1}\boldsymbol{\alpha}^{n-k+1}\right),\ldots,\boldsymbol{u}^{2}\star\left(\boldsymbol{\alpha}^{\color{red}{2n-2k-\ell-3}}-\varTheta_{n-k-1}\boldsymbol{\alpha}^{2n-2k-\ell-2}\right),$$
$$\boldsymbol{u}^{2}\star\left(\boldsymbol{\alpha}^{2n-2k-2\ell-3}-\varTheta_{n-k-\ell-1}\boldsymbol{\alpha}^{\color{red}{2n-2k-\ell-2}}\right),$$
$$\boldsymbol{u}^{2}\star\left(\Gamma_{n-k-\ell-1,s_2}-\varTheta_{s_2}\boldsymbol{\alpha}^{\color{red}{2n-2k-\ell-1}}\right),\ldots,\boldsymbol{u}^{2}\star\left(\Gamma_{n-k-1,s_2}-\varTheta_{s_2}\boldsymbol{\alpha}^{\color{red}{2n-2k-1}}\right)$$ are $\mathbb{F}_{q}$-linearly independent, where $$\Gamma_{s_1,s_2}=\boldsymbol{\alpha}^{s_1+s_2}-\varTheta_{s_1}\boldsymbol{\alpha}^{n-k+s_2}+\varTheta_{s_1}\varTheta_{s_2}\boldsymbol{\alpha}^{2n-2k}\left\{s_1,s_2\in\left\{n-k-\ell-1,\ldots,n-k-1 \right\}\right\}.$$ Furthermore, 
$$\dim\left(\left(\mathcal{C}^{\perp}\left(\boldsymbol{\alpha},\boldsymbol{1},\boldsymbol{\eta}\right)\right)^2\right)\geq 2n-2k,$$
thus by Lemma \ref{GRScodeschur2}, the code $\mathcal{C}_{k-1}\left(\boldsymbol{\alpha},\boldsymbol{v},\boldsymbol{\eta}\right)$ is non-GRS.

Secondly, for $1+\sum\limits_{t=0}^{\ell}\eta_{t}S_{t+1}=0$ and $\ell\geq 1$,  by Definition \ref{schurproduct},  we have
$$
\begin{aligned}
	&\left(\mathcal{C}^{\perp}\left(\boldsymbol{\alpha},\boldsymbol{1},\boldsymbol{\eta}\right)\right)^2\\
	=&\left\langle\boldsymbol{u}\star\boldsymbol{\alpha}^{i},\boldsymbol{u}\star\left(\boldsymbol{\alpha}^{t_1}-\Omega_{t_1}\boldsymbol{\alpha}^{n-k-(\ell+1) }\right),\boldsymbol{u}\star\boldsymbol{\alpha}^{n-k}\right\rangle\star\left\langle\boldsymbol{u}\star\boldsymbol{\alpha}^{j},\boldsymbol{u}\star\left(\boldsymbol{\alpha}^{t_2}-\Omega_{t_2}\boldsymbol{\alpha}^{n-k-(\ell+1) }\right),\boldsymbol{u}\star\boldsymbol{\alpha}^{n-k}\right\rangle\\
	&\left(i,j\in\left\{0,1,\ldots,n-k-\ell-2\right\},t_1,t_2\in\left\{n-k-\ell,\ldots,n-k-1\right\}\right)\\
  =&\left\langle\boldsymbol{u}^2\star\boldsymbol{\alpha}^{i+j},\boldsymbol{u}^2\star\boldsymbol{\alpha}^{i}\star\left(\boldsymbol{\alpha}^{t_2}-\Omega_{t_2}\boldsymbol{\alpha}^{n-k-(\ell+1) }\right),\boldsymbol{u}^2\star\boldsymbol{\alpha}^{n-k+i},\boldsymbol{u}^2\star\left(\boldsymbol{\alpha}^{t_1}-\Omega_{t_1}\boldsymbol{\alpha}^{n-k-(\ell+1) }\right)\star\boldsymbol{\alpha}^{j},\right.\\
  &\left.\boldsymbol{u}^2\star\left(\boldsymbol{\alpha}^{t_1}-\Omega_{t_1}\boldsymbol{\alpha}^{n-k-(\ell+1) }\right)\star\left(\boldsymbol{\alpha}^{t_2}-\Omega_{t_2}\boldsymbol{\alpha}^{n-k-(\ell+1) }\right),\boldsymbol{u}^2\star\left(\boldsymbol{\alpha}^{t_1}-\Omega_{t_1}\boldsymbol{\alpha}^{n-k-(\ell+1) }\right)\star\boldsymbol{\alpha}^{n-k},\right.\\
  &\left.\boldsymbol{u}^2\star\boldsymbol{\alpha}^{n-k+j},\boldsymbol{u}^2\star\boldsymbol{\alpha}^{n-k}\star\left(\boldsymbol{\alpha}^{t_2}-\Omega_{t_2}\boldsymbol{\alpha}^{n-k-(\ell+1) }\right),\boldsymbol{u}^2\star\boldsymbol{\alpha}^{2n-2k}\right\rangle\\
  	&\left(i,j\in\left\{0,1,\ldots,n-k-\ell-2\right\},t_1,t_2\in\left\{n-k-\ell,\ldots,n-k-1\right\}\right)\\
  	=&\left\langle\boldsymbol{u}^2\star\boldsymbol{\alpha}^{i+j},\boldsymbol{u}^2\star\left(\boldsymbol{\alpha}^{i+t_2}-\Omega_{t_2}\boldsymbol{\alpha}^{n-k-\ell-1+i }\right),\boldsymbol{u}^2\star\boldsymbol{\alpha}^{n-k+i},\boldsymbol{u}^2\star\left(\boldsymbol{\alpha}^{t_1+j}-\Omega_{t_1}\boldsymbol{\alpha}^{n-k-\ell-1+j }\right),\right.\\
  	&\left.\boldsymbol{u}^2\star\left(\boldsymbol{\alpha}^{t_1+t_2}-\Omega_{t_1}\boldsymbol{\alpha}^{n-k-\ell-1+t_2 }-\Omega_{t_2}\boldsymbol{\alpha}^{n-k-\ell-1+t_1 }+\Omega_{t_1}\Omega_{t_2}\boldsymbol{\alpha}^{2n-2k-2\ell-2 }\right),\right.\\
  	&\left.\boldsymbol{u}^2\star\left(\boldsymbol{\alpha}^{n-k+t_1}-\Omega_{t_1}\boldsymbol{\alpha}^{2n-2k-\ell-1 }\right),\boldsymbol{u}^2\star\boldsymbol{\alpha}^{n-k+j},\boldsymbol{u}^2\star\left(\boldsymbol{\alpha}^{n-k+t_2}-\Omega_{t_2}\boldsymbol{\alpha}^{2n-2k-\ell-1 }\right),\boldsymbol{u}^2\star\boldsymbol{\alpha}^{2n-2k}\right\rangle\\
  	&\left(i,j\in\left\{0,1,\ldots,n-k-\ell-2\right\},t_1,t_2\in\left\{n-k-\ell,\ldots,n-k-1\right\}\right).
\\  	
\end{aligned}
$$
By $2k>n\geq k+\ell+2$, we have $$n-k-\ell-2\leq 2n-2k-2\ell-4$$ and $$2n-2k-1<n-1,$$ then 
$$\boldsymbol{u}^{2}\star\boldsymbol{\alpha}^{\color{red}{0}},\boldsymbol{u}^{2}\star\boldsymbol{\alpha}^{\color{red}{1}},\ldots\boldsymbol{u}^{2}\star\boldsymbol{\alpha}^{\color{red}{n-k-\ell-2}},$$
$$\boldsymbol{u}^2\star\left(\boldsymbol{\alpha}^{n-k-1}-\Omega_{n-k-1}\boldsymbol{\alpha}^{\color{red}{n-k-\ell-1}}\right),\ldots,\boldsymbol{u}^2\star\left(\boldsymbol{\alpha}^{2n-2k-\ell-3}-\Omega_{n-k-1}\boldsymbol{\alpha}^{\color{red}{2n-2k-2\ell-3}}\right),$$
$$\boldsymbol{u}^2\star\left(\boldsymbol{\alpha}^{\color{red}{2n-2k-2\ell-2}}-\Omega_{n-k-\ell}\boldsymbol{\alpha}^{2n-2k-2\ell-3}\right),\ldots,\boldsymbol{u}^2\star\left(\boldsymbol{\alpha}^{\color{red}{2n-2k-\ell-4}}-\Omega_{n-k-2}\boldsymbol{\alpha}^{2n-2k-2\ell-3}\right),$$
$$
\boldsymbol{u}^2\star\boldsymbol{\alpha}^{\color{red}{2n-2k-\ell-3}},\boldsymbol{u}^2\star\boldsymbol{\alpha}^{\color{red}{2n-2k-\ell-2}},$$
$$\boldsymbol{u}^2\star\left(\boldsymbol{\alpha}^{\color{red}{2n-2k-\ell-1}}-\Omega_{n-k-\ell}\boldsymbol{\alpha}^{2n-2k-\ell-2 }-\Omega_{n-k-1}\boldsymbol{\alpha}^{2n-2k-2\ell-1 }-\Omega_{n-k-\ell}\Omega_{n-k-1}\boldsymbol{\alpha}^{2n-2k-2\ell-2 }\right),
$$
$$\boldsymbol{u}^2\star\left(\boldsymbol{\alpha}^{\color{red}{2n-2k-\ell}}-\Omega_{n-k-\ell}\boldsymbol{\alpha}^{2n-2k-\ell-1}\right),\ldots,\boldsymbol{u}^2\star\left(\boldsymbol{\alpha}^{\color{red}{2n-2k-1}}-\Omega_{n-k-1}\boldsymbol{\alpha}^{2n-2k-\ell-1}\right)$$
are $\mathbb{F}_{q}$-linearly independent,  furthermore, 
$$\dim\left(\left(\mathcal{C}_{k-1}^{\perp}\left(\boldsymbol{\alpha},\boldsymbol{1},\boldsymbol{\eta}\right)\right)^2\right)\geq 2n-2k,$$
thus by Lemma \ref{GRScodeschur2}, the code $\mathcal{C}_{k-1}\left(\boldsymbol{\alpha},\boldsymbol{v},\boldsymbol{\eta}\right)$ is non-GRS. 

From the above discussions, we complete the proof of Theorem $\ref{nonRS2kn}$. 
\section{Some examples}\label{sec6}
In this section, for Theorems \ref{l=0yes}-\ref{l=1n=2k+1yes2}, Corollary \ref{lgeq1n=2k+l+1yes2}, Theorem \ref{lgeq1ngeq2k+2l+2yes} and Theorem \ref{LCDMDS}, we give the corresponding example, respectively, which show that there exist many self-orthogonal codes and LCD MDS codes.
\begin{example}
	Let $\left(q,n,k,\ell\right)=\left(17,11,5,1\right)$, $\boldsymbol{\alpha}=\left(1, 2, 3, 4, 5, 6, 7, 9, 10, 15, 16\right)$ and $$\small\begin{aligned}\boldsymbol{\eta}\in&\left\{(9, 9), (10, 14), (3, 7), (8, 10), (5, 1), (10, 7), (15, 16), (5, 16), (16, 5), (8, 6),(15, 9), (16, 14)\right\} \in\mathbb{F}_{17}^{2}\backslash\left\{\boldsymbol{0}\right\} 
	\end{aligned}.$$ 
	By directly calculating, we have
	$$\boldsymbol{u}=\left(15, 15, 4, 9, 4, 15, 15, 8, 13, 13, 8\right),$$
	$$\boldsymbol{v}=\left(1, 1, 7, 2, 7, 1, 1, 8, 6, 6, 8\right)$$
	and there exists $\lambda=8$ such that $\lambda\boldsymbol{u}=\boldsymbol{v}^2$. Thus, the code $\mathcal{C}_{k-1}\left(\boldsymbol{\alpha},\boldsymbol{v},\boldsymbol{\eta}\right)$ has the following generator matrix 
	$$\footnotesize\begin{aligned}
	\boldsymbol{G}_{k,+}\in&\left\{	\begin{pmatrix}
		1 & 1 & 7 & 2 & 7 & 1 & 1 & 8 & 6 & 6 & 8 \\
		1 & 2 & 4 & 8 & 1 & 6 & 7 & 4 & 9 & 5 & 9 \\
		1 & 4 & 12 & 15 & 5 & 2 & 15 & 2 & 5 & 7 & 8 \\
		1 & 8 & 2 & 9 & 8 & 12 & 3 & 1 & 16 & 3 & 9 \\
		2 & 13 & 8 & 5 & 11 & 3 & 14 & 6 & 1 & 5 & 8
	\end{pmatrix},\begin{pmatrix}
	1 & 1 & 7 & 2 & 7 & 1 & 1 & 8 & 6 & 6 & 8 \\
	1 & 2 & 4 & 8 & 1 & 6 & 7 & 4 & 9 & 5 & 9 \\
	1 & 4 & 12 & 15 & 5 & 2 & 15 & 2 & 5 & 7 & 8 \\
	1 & 8 & 2 & 9 & 8 & 12 & 3 & 1 & 16 & 3 & 9 \\
	8&8&7&3&9&16&2&9& 1 &16&6
	\end{pmatrix}\right.\\
	&\left.\begin{pmatrix}
	1 & 1 & 7 & 2 & 7 & 1 & 1 & 8 & 6 & 6 & 8 \\
	1 & 2 & 4 & 8 & 1 & 6 & 7 & 4 & 9 & 5 & 9 \\
	1 & 4 & 12 & 15 & 5 & 2 & 15 & 2 & 5 & 7 & 8 \\
	1 & 8 & 2 & 9 & 8 & 12 & 3 & 1 & 16 & 3 & 9 \\
	11&16&13&12&7&13&15&0&0&15&6
	\end{pmatrix},\begin{pmatrix}
	1 & 1 & 7 & 2 & 7 & 1 & 1 & 8 & 6 & 6 & 8 \\
	1 & 2 & 4 & 8 & 1 & 6 & 7 & 4 & 9 & 5 & 9 \\
	1 & 4 & 12 & 15 & 5 & 2 & 15 & 2 & 5 & 7 & 8 \\
	1 & 8 & 2 & 9 & 8 & 12 & 3 & 1 & 16 & 3 & 9 \\
	2&11&10&12&12&4&12&8&2&3&7
	\end{pmatrix},\cdots\right\}, 
\end{aligned}$$
	then by Theorem \ref{l=1n=2k+1yes1}, the code $\mathcal{C}_{k-1}\left(\boldsymbol{\alpha},\boldsymbol{v},\boldsymbol{\eta}\right)$ generalized by the above $\boldsymbol{G}_{k,+}^{(\ell)}$ is self-orthogonal. In fact, based on the Magma programe, the code $\mathcal{C}_{k-1}\left(\boldsymbol{\alpha},\boldsymbol{v},\boldsymbol{\eta}\right)$ is a self-orthogonal code with the parameters $\left\{[11,5,5]_{17},[11,5,5]_{17},[11,5,5]_{17},[11,5,5]_{17},\cdots\right\}$.
\end{example}
\begin{example}
	Let $\left(q,n,k,\ell\right)=\left(31,13,6,1\right)$,$\boldsymbol{\alpha}=\left(1, 2, 3, 4, 5, 6, 7, 9, 10, 13, 24, 25, 28\right),$ and
	$$\boldsymbol{\eta}\in\left\{\left(30,13\right),\left(26,28\right)\right\} \in\mathbb{F}_{31}^{2}\backslash\left\{\boldsymbol{0}\right\}.$$ 
	By directly calculating, we have
	$$\boldsymbol{u}=\left(28, 5, 5, 4, 2, 20, 25, 4, 19, 5, 20, 2, 16\right),$$
	$$\boldsymbol{v}=\left(1, 9, 9, 28, 19, 18, 8, 28, 2, 9, 18, 19, 25\right)$$
	and there exists $\lambda=10$ such that $\lambda\boldsymbol{u}=\boldsymbol{v}^2$. Thus, the code $\mathcal{C}_{k-1}\left(\boldsymbol{\alpha},\boldsymbol{v},\boldsymbol{\eta}\right)$ has the following generator matrix 
	$$\begin{aligned}
	\boldsymbol{G}_{k,+}\in&\left\{\begin{pmatrix}
		1 & 9 & 9 & 28 & 19 & 18 & 8 & 28 & 2 & 9 & 18 & 19 & 25 \\
		1 & 18 & 27 & 19 & 2 & 15 & 25 & 4 & 20 & 24 & 29 & 10 & 18 \\
		1 & 5 & 19 & 14 & 10 & 28 & 20 & 5 & 14 & 2 & 14 & 2 & 8 \\
		1 & 10 & 26 & 25 & 19 & 13 & 16 & 14 & 16 & 26 & 26 & 19 & 7 \\
		1 & 20 & 16 & 7 & 2 & 16 & 19 & 2 & 5 & 28 & 4 & 10 & 10 \\
		13 & 25 & 2 & 5 & 17 & 25 & 6 & 24 & 8 & 4 & 13 & 20 & 28 \\
	\end{pmatrix},\right.\\
	&\left.\begin{pmatrix}
	1 & 9 & 9 & 28 & 19 & 18 & 8 & 28 & 2 & 9 & 18 & 19 & 25 \\
	1 & 18 & 27 & 19 & 2 & 15 & 25 & 4 & 20 & 24 & 29 & 10 & 18 \\
	1 & 5 & 19 & 14 & 10 & 28 & 20 & 5 & 14 & 2 & 14 & 2 & 8 \\
	1 & 10 & 26 & 25 & 19 & 13 & 16 & 14 & 16 & 26 & 26 & 19 & 7 \\
	1 & 20 & 16 & 7 & 2 & 16 & 19 & 2 & 5 & 28 & 4 & 10 & 10 \\
	24&28&16&15&2&23&14&11&3&11&8&1&20 \\
	\end{pmatrix}\right\},
\end{aligned}$$
	then by Theorem \ref{l=1n=2k+1yes2}, the code $\mathcal{C}_{k-1}\left(\boldsymbol{\alpha},\boldsymbol{v},\boldsymbol{\eta}\right)$ generalized by the above 	$\boldsymbol{G}_{k,+}^{(\ell)}$ is self-orthogonal. In fact, based on the Magma programe, the code $\mathcal{C}_{k-1}\left(\boldsymbol{\alpha},\boldsymbol{v},\boldsymbol{\eta}\right)$ is a self-orthogonal code with the parameters $[13,6,6]_{31}$.
\end{example}
\begin{example}
	Let $\left(q,n,k,\ell,\mu\right)=\left(37,18,7,3,36\right)$ and $\boldsymbol{\eta}=\left(1,4,7,9\right) \in\mathbb{F}_{37}^{4}\backslash\left\{\boldsymbol{0}\right\}$. It's easy to know that $$\mathrm{ord}\left(\mu\right)=\mathrm{ord}\left(36\right)=2\mid \frac{q-1}{n}=2.$$ And based on the Magma programe, we know
	$$\begin{aligned}
		x^{18}-36=&(x-2)(x-5)(x-6)(x-8)(x-13)(x-14)(x-15)(x-17)(x-18)(x-19)\\
		&(x-20)(x-22)(x-23)(x-24)(x-29)(x-31)(x-32)(x-35). 
	\end{aligned}$$
	Furthermore, by taking
	$$\boldsymbol{\alpha}=\left(2, 5, 6, 8, 13, 14, 15, 17, 18, 19, 20, 22, 23, 24, 29, 31, 32, 35\right)\in\mathbb{F}_{37}^{18},$$
	and directly calculating, we have $$
	\boldsymbol{u}=\left(4, 10, 12, 16, 26, 28, 30, 34, 36, 1, 3, 7, 9, 11, 21, 25, 27, 33\right),$$
	$$\boldsymbol{v}=\left(2, 11, 7, 4, 10, 18, 17, 16, 6, 1, 15, 9, 3, 14, 13, 5, 8, 12\right)\in\mathbb{F}_{37}^{18},$$ 
	and there exists $\lambda=1$ such that $\lambda\boldsymbol{u}=\boldsymbol{v}^2$.
	Thus, the code $\mathcal{C}_{k-1}\left(\boldsymbol{\alpha},\boldsymbol{v},\boldsymbol{\eta}\right)$ has the following generator matrix 
	\begin{equation}\label{example(1)}
		\small\boldsymbol{G}_{k,+}=\begin{pmatrix}
			2 & 11 & 7 & 4 & 10 & 18 & 17 & 16 & 6 & 1 & 15 & 9 & 3 & 14 & 13 & 5 & 8 & 12 \\
			4 & 18 & 5 & 32 & 19 & 30 & 33 & 13 & 34 & 19 & 4 & 13 & 32 & 3 & 7 & 7 & 34 & 13 \\
			8 & 16 & 30 & 34 & 25 & 13 & 14 & 36 & 20 & 28 & 6 & 27 & 33 & 35 & 18 & 32 & 15 & 11 \\
			16 & 6 & 32 & 13 & 29 & 34 & 25 & 20 & 27 & 14 & 9 & 2 & 19 & 26 & 4 & 30 & 36 & 15 \\
			32 & 30 & 7 & 30 & 7 & 32 & 5 & 7 & 5 & 7 & 32 & 7 & 30 & 32 & 5 & 5 & 5 & 7 \\
			27 & 2 & 5 & 18 & 17 & 4 & 1 & 8 & 16 & 22 & 11 & 6 & 24 & 28 & 34 & 7 & 12 & 23 \\
			23 & 15 & 25 & 22 & 16 & 3 & 33 & 14 & 9 & 4 & 19 & 2 & 22 & 15 & 20 & 12 & 23 & 35
		\end{pmatrix},
	\end{equation}
	then by Corollary \ref{lgeq1n=2k+l+1yes2}, the code $\mathcal{C}_{k-1}\left(\boldsymbol{\alpha},\boldsymbol{v},\boldsymbol{\eta}\right)$ generalized by
	$(\ref{example(1)})$ is self-orthogonal. In fact, based on the Magma programe, the code $\mathcal{C}_{k-1}\left(\boldsymbol{\alpha},\boldsymbol{v},\boldsymbol{\eta}\right)$ is a self-orthogonal code with the parameters $[18,7,10]_{37}$. 
\end{example}
\begin{example}
	Let $\left(q,n,k,\ell\right)=\left(31,16,5,2\right)$, $\boldsymbol{\eta}=\left(1,2,3\right) \in\mathbb{F}_{31}^{3}\backslash\left\{\boldsymbol{0}\right\}$ and 
	$$\boldsymbol{\alpha}=\left(1, 2, 3, 4, 5, 6, 7, 8, 9, 10, 11, 13, 18, 25, 27, 29\right).$$ 
	By directly calculating, we have
	$$\boldsymbol{u}=\left(15, 13, 27, 29, 27, 17, 6, 6, 24, 26, 26, 22, 22, 3, 3, 13\right),$$
	$$\boldsymbol{v}=\left(1,25,16,2,16,20,9,9,18,14,14,7,7,5,5,25\right)$$
	and there exists $\lambda=29$ such that $\lambda\boldsymbol{u}=\boldsymbol{v}^2$. Thus, the code $\mathcal{C}_{k-1}\left(\boldsymbol{\alpha},\boldsymbol{v},\boldsymbol{\eta}\right)$ has the following generator matrix 
	\begin{equation}\label{example(2)}
		\begin{pmatrix}
1&25&16&2&16&20&9&9&18&14&14&7&7&5&5&25\\
1&19&17&8&18&27&1&10&7&16&30&29&2&1&11&12\\
1&7&20&1&28&7&7&18&1&5&20&5&5&25&18&7\\
1&14&29&4&16&11&18&20&9&19&3&3&28&5&21& 17&\\
7&19&2&6&8&25&7&26&26&10&0&23&7&8&21&20
		\end{pmatrix},
	\end{equation}
	then by Theorem \ref{lgeq1ngeq2k+2l+2yes}, the code $\mathcal{C}_{k-1}\left(\boldsymbol{\alpha},\boldsymbol{v},\boldsymbol{\eta}\right)$ generalized by
	$(\ref{example(2)})$ is self-orthogonal . In fact, based on the Magma programe, the code $\mathcal{C}_{k-1}\left(\boldsymbol{\alpha},\boldsymbol{v},\boldsymbol{\eta}\right)$ is a self-orthogonal code with the parameters $[16,5,9]_{31}$.
\end{example}

\begin{example}
	Let $\left(q,n,k,\ell\right)=\left(31,16,4,0\right)$, $\boldsymbol{\eta}=1 \in\mathbb{F}_{31}^{*}$ and 
	$$\boldsymbol{\alpha}=\left(1, 2, 3, 4, 5, 6, 7, 8, 9, 10, 11, 13, 18, 25, 27, 29\right).$$ 
	By directly calculating, we have
	$$\boldsymbol{u}=\left(15, 13, 27, 29, 27, 17, 6, 6, 24, 26, 26, 22, 22, 3, 3, 13\right),$$
	$$\boldsymbol{v}=\left(1,25,16,2,16,20,9,9,18,14,14,7,7,5,5,25\right)$$
	and there exists $\lambda=29$ such that $\lambda\boldsymbol{u}=\boldsymbol{v}^2$. Thus, the code $\mathcal{C}_{k-1}\left(\boldsymbol{\alpha},\boldsymbol{v},\boldsymbol{\eta}\right)$ has the following generator matrix 
	\begin{equation}\label{example(3)}
		\begin{pmatrix}
			1&25&16&2&16&20&9&9&18&14&14&7&7&5&5&25\\
			1&19&17&8&18&27&1&10&7&16&30&29&2&1&11&12\\
			1&7&20&1&28&7&7&18&1&5&20&5&5&25&18&7\\ 
			2&11&23&20&3&15&20&25&28&23&5&11&5&6&30&14
		\end{pmatrix},
	\end{equation}
	then by Theorem \ref{l=0ngeq2k+2}, the code $\mathcal{C}_{k-1}\left(\boldsymbol{\alpha},\boldsymbol{v},\boldsymbol{\eta}\right)$ generalized by
	$(\ref{example(3)})$ is self-orthogonal . In fact, based on the Magma programe, the code $\mathcal{C}_{k-1}\left(\boldsymbol{\alpha},\boldsymbol{v},\boldsymbol{\eta}\right)$ is a self-orthogonal code with the parameters $[16,4,12]_{31}$.
\end{example} 

\begin{example}
	Let $\left(q,n,k,\ell\right)=\left(17,8,5,2\right),$ 
	$$\begin{aligned}
	\boldsymbol{\alpha}\in&\left\{\left(7, 11, 12, 16, 8, 6, 3, 1\right),\left(7, 3, 10, 6, 9, 1, 11, 4\right),\left(5, 13, 4, 6, 11, 2, 10, 12\right),\right.\\
	&\left.\left(3, 4, 9, 5, 11, 1, 8, 16\right),\left(10, 15, 1, 2, 16, 3, 11, 14\right),\left(7, 16, 10, 13, 5, 11, 12, 2\right),\cdots \right\}, 
	\end{aligned}$$
	$$\begin{aligned}
	\boldsymbol{v}\in&\left\{\left(6, 11, 16, 9, 15, 12, 9, 9\right),\left(13, 14, 15, 5, 6, 15, 4, 7\right),\left(9, 12, 7, 11, 7, 12, 6, 10\right),\right.\\
	&\left.\left(16, 5, 12, 2, 16, 13, 11, 3\right) ,\left(1, 9, 13, 13, 5, 5, 13, 14\right),\left( 3, 3, 10, 4, 12, 12, 4, 8\right),\cdots\right\},	
	\end{aligned}$$ and 
	$$ \boldsymbol{\eta}\in\left\{\left(16, 8, 4\right),\left(1, 8, 10\right),\left(1, 12, 13\right),\left(6, 3, 5\right),\left(13, 16, 12\right),\left(16, 12, 16\right),\cdots\right\}.$$
Thus, the code $\mathcal{C}_{k-1}\left(\boldsymbol{\alpha},\boldsymbol{v},\boldsymbol{\eta}\right)$ has the following generator matrix
	$$
	\begin{aligned}
	\boldsymbol{G}_{k,+}\in&\left\{\begin{pmatrix}
		6 & 11 & 16 & 9 & 15 & 12 & 9 & 9 \\
		8 & 2 & 5 & 8 & 1 & 4 & 10 & 9 \\
		5 & 5 & 9 & 9 & 8 & 7 & 13 & 9 \\
		1 & 4 & 6 & 8 & 13 & 8 & 5 & 9 \\
		15 & 5 & 14 & 3 & 6 & 10 & 1 & 6
	\end{pmatrix},\begin{pmatrix}
	6 & 11 & 16 & 9 & 15 & 12 & 9 & 9 \\
	6 & 8 & 14 & 13 & 3 & 15 & 10 & 11 \\
	8 & 7 & 4 & 10 & 10 & 15 & 8 & 10 \\
	5 & 4 & 6 & 9 & 5 & 15 & 3 & 6 \\
	5 & 4 & 8 & 4 & 14 & 11 & 7 & 5
	\end{pmatrix},\right.\\
	&\left.\begin{pmatrix}
		16 & 5 & 12 & 2 & 16 & 13 & 11 & 3 \\
		14 & 3 & 6 & 10 & 6 & 13 & 3 & 14 \\
		8 & 12 & 3 & 16 & 15 & 13 & 7 & 3 \\
		7 & 14 & 10 & 12 & 12 & 13 & 5 & 14 \\
		10 & 10 & 12 & 0 & 16 & 8 & 10 & 13
	\end{pmatrix},\begin{pmatrix}
	10 & 16 & 6 & 6 & 16 & 3 & 14 & 5 \\
	16 & 4 & 7 & 2 & 6 & 6 & 4 & 9 \\
	12 & 1 & 11 & 12 & 15 & 12 & 6 & 6 \\
	9 & 13 & 10 & 4 & 12 & 7 & 9 & 4 \\
	11 & 6 & 3 & 11 & 16 & 1 & 3 & 4
	\end{pmatrix},\right.\\
	&\left.\begin{pmatrix}
		1 & 9 & 13 & 13 & 5 & 5 & 13 & 14 \\
		10 & 16 & 13 & 9 & 12 & 15 & 7 & 9 \\
		15 & 2 & 13 & 1 & 5 & 11 & 9 & 7 \\
		14 & 13 & 13 & 2 & 12 & 16 & 14 & 13 \\
		14 & 3 & 2 & 0 & 11 & 6 & 15 & 2
	\end{pmatrix},\begin{pmatrix}
		3 & 3 & 10 & 4 & 12 & 12 & 4 & 8 \\
		4 & 14 & 15 & 1 & 9 & 13 & 14 & 16 \\
		11 & 3 & 14 & 13 & 11 & 7 & 15 & 15 \\
		9 & 14 & 4 & 16 & 4 & 9 & 10 & 13 \\
		12 & 11 & 7 & 7 & 3 & 7 & 6 & 11
	\end{pmatrix},\cdots\right\}.
	\end{aligned}
	$$
	Based on the Magma programe, the code $\mathcal{C}_{k-1}\left(\boldsymbol{\alpha},\boldsymbol{v},\boldsymbol{\eta}\right)$ generalized by the above matrix $\boldsymbol{G}_{k,+}^{(\ell)}$  is NMDS with the parameters $[8,5,3]_{17}$.
\end{example}

\begin{example}
Let $(q,n,k,\ell)=\left(43,9,4,1\right)$,
$$\boldsymbol{\eta}\in\left\{\left(27, 18\right),\left(18,37\right),\left(37,12\right),\left(23,4\right)\right\},$$
and
$$
\begin{aligned}
\boldsymbol{\alpha}\in&\left\{\left(1, 2, 3, 4, 5, 6, 7, 18, 33\right),\left(1, 2, 3, 4, 5, 6, 8, 13, 20\right),\right.\\
&\left.\left(1, 2, 3, 4, 5, 6, 8, 14, 17\right),\left(1, 2, 3, 4, 5, 6, 8, 27, 28\right)\right\}.
\end{aligned}$$
By directly calculating, we have
$$
\begin{aligned}
\boldsymbol{u}\in&\left\{\left(6, 16, 40, 11, 40, 16, 6, 40, 40\right),\left(21, 14, 24, 4, 14, 1, 9, 1, 41\right),\right.\\
&\left.\left(4, 24, 17, 17, 6, 6, 10, 31, 14\right),\left(41, 4, 24, 16, 4, 38, 15, 17, 13\right)\right\}
\end{aligned},$$ 
$$
\begin{aligned}
\boldsymbol{v}\in&\left\{\left(1, 24, 35, 40, 35, 24, 1, 35, 35\right),\left(1, 31, 9, 11, 31, 16, 38, 16, 41\right),\right.\\
&\left.\left(1, 36, 31, 31, 25, 25, 14, 13, 38\right),\left(42, 27, 26, 32, 27, 29, 33, 20, 12\right)\right\}
\end{aligned},$$
and there exists $\lambda\in\left\{36,41,11,21\right\}$ such that $\lambda\boldsymbol{u}=\boldsymbol{v}^2$. Thus, the code $\mathcal{C}_{k-1}\left(\boldsymbol{\alpha},\boldsymbol{v},\boldsymbol{\eta}\right)$ has the following generator matrix
$$
\begin{aligned}
\boldsymbol{G}_{k,+}\in&\left\{\begin{pmatrix}
	1 & 24 & 35 & 40 & 35 & 24 & 1 & 35 & 35 \\
	1 & 5  & 19 & 31 & 3  & 15 & 7 & 28 & 37 \\
	1 & 10 & 14 & 38 & 15 & 4  & 6 & 31 & 17 \\
	3 & 3  & 14 & 15 & 4  & 28 & 3 & 2  & 9  \\
\end{pmatrix},\begin{pmatrix}
1  & 31 & 9  & 11 & 31 & 16 & 38 & 16 & 41 \\
1  & 19 & 27 & 1  & 26 & 10 & 3  & 36 & 3  \\
1  & 38 & 38 & 4  & 1  & 17 & 24 & 38 & 17 \\
13 & 42 & 28 & 19 & 6  & 8  & 36 & 24 & 29 \\
\end{pmatrix}\right.\\
&\left.\begin{pmatrix}
	1  & 36 & 31 & 31 & 25 & 25 & 14 & 13 & 38 \\
	1  & 29 & 7  & 38 & 39 & 21 & 26 & 10 & 1  \\
	1  & 15 & 21 & 23 & 23 & 40 & 36 & 11 & 17 \\
	7  & 35 & 14 & 25 & 33 & 35 & 1  & 8  & 16 \\
\end{pmatrix},\begin{pmatrix}
42 & 27 & 26 & 32 & 27 & 29 & 33 & 20 & 12 \\
42 & 11 & 35 & 42 & 6  & 2  & 6  & 24 & 35 \\
42 & 22 & 19 & 39 & 30 & 12 & 5  & 3  & 34 \\
15 & 20 & 22 & 25 & 21 & 37 & 10 & 26 & 25 \\
\end{pmatrix}\right\}
,\end{aligned}
$$
then by Theorem \ref{LCDMDS}, for any $\beta\in\mathbb{F}_{43}\backslash\left\{-1,1\right\}$,  the code $\mathcal{C}\left(\boldsymbol{\alpha},\boldsymbol{v}^{\prime},\boldsymbol{\eta}\right)$ generalized by
$$
\footnotesize\begin{aligned}
	\boldsymbol{G}_{k,+}^{\prime}\in&\left\{\begin{pmatrix}
		1 & 24 & 35 & 40 & 35\beta & 24\beta & 1\beta & 35\beta & 35\beta \\
		1 & 5  & 19 & 31 & 3\beta  & 15\beta & 7\beta & 28\beta & 37\beta \\
		1 & 10 & 14 & 38 & 15\beta & 4\beta  & 6\beta & 31\beta & 17\beta \\
		3 & 3  & 14 & 15 & 4\beta  & 28\beta & 3\beta & 2\beta  & 9\beta  \\
	\end{pmatrix},\begin{pmatrix}
		1  & 31 & 9  & 11 & 31\beta & 16\beta & 38\beta & 16\beta & 41\beta \\
		1  & 19 & 27 & 1  & 26\beta & 10\beta & 3\beta  & 36\beta & 3\beta  \\
		1  & 38 & 38 & 4  & \beta  & 17\beta & 24\beta & 38\beta & 17\beta \\
		13 & 42 & 28 & 19 & 6\beta  & 8\beta  & 36\beta & 24\beta & 29\beta \\
	\end{pmatrix}\right.\\
	&\left.\begin{pmatrix}
		1  & 36 & 31 & 31 & 25\beta & 25\beta & 14\beta & 13\beta & 38\beta \\
		1  & 29 & 7  & 38 & 39\beta & 21\beta & 26\beta & 10\beta & \beta  \\
		1  & 15 & 21 & 23 & 23\beta & 40\beta & 36\beta & 11\beta & 17\beta \\
		7  & 35 & 14 & 25 & 33\beta & 35\beta & \beta  & 8\beta  & 16\beta \\
	\end{pmatrix},\begin{pmatrix}
		42 & 27 & 26 & 32 & 27\beta & 29\beta & 33\beta & 20\beta & 12\beta \\
		42 & 11 & 35 & 42 & 6\beta  & 2\beta  & 6\beta  & 24\beta & 35\beta \\
		42 & 22 & 19 & 39 & 30\beta & 12\beta & 5\beta  & 3\beta  & 34\beta \\
		15 & 20 & 22 & 25 & 21\beta & 37\beta & 10\beta & 26\beta & 25\beta \\
	\end{pmatrix}\right\}
\end{aligned}
$$
is LCD MDS. In fact, based on the Magma programe, the code $\mathcal{C}\left(\boldsymbol{\alpha},\boldsymbol{v}^{\prime},\boldsymbol{\eta}\right)$ generalized by the above $ \boldsymbol{G}_{k,+}^{\prime}$ is LCD MDS with the parameters $[9,4,6]_{43}$.
\end{example}
\section{Conclusions}\label{sec7}
In this paper, we presents a comprehensive analysis of a generalized class of $(\mathcal{L}, \mathcal{P})$ TGRS codes with $\ell+1$ twists. By taking an explicit form of their parity-check matrices, we successfully characterize their self-orthogonal and NMDS properties. A significant contribution of this work is the partial resolution of the open problem posed by Hu et al., which bridges the gap toward a full characterization of twisted algebraic codes. Furthermore, its results expand the library of LCD MDS codes, providing designers with greater flexibility in selecting code parameters for practical implementation. Utilizing the Schur product method, we rigorously confirms the non-GRS nature of these codes, particularly under the improved condition $2k>n$. These findings not only advance the theoretical understanding of twisted code structures but also lay the groundwork for future investigations into their Schur square dimensions and potential applications in code-based cryptography and distributed storage systems. The main results of this paper are shown in the following table.
\begin{threeparttable}[H]
	\centering
	\footnotesize
	\caption{The existence of the self-orthogonal code $\mathcal{C}_{k-1}\left(\boldsymbol{\alpha},\boldsymbol{v},\boldsymbol{\eta}\right)$}  
	\label{+TGRScodeselfdual}
	\begin{tabular}{|c|c|c|c|c|c|c|c|}
		\hline 
		$\ell$&$n$&$\sum\limits_{t=0}^{n}\alpha_{t}$&$1+\sum\limits_{t=0}^{\ell}\eta_{t}S_{t+1}$&$\mathrm{Char}\left(\mathbb{F}_{q}\right)$&$\boldsymbol{\eta}$&self-orthogonal&reference\\
		\hline
		\multirow{8}{*}{$0$}&\multirow{5}{*}{$2k$}&\multirow{2}{*}{$\neq 0$}&\multirow{2}{*}{$\neq 0$}&$=2$ & &$\times$&Theorem \ref{l=0not} (2)\\
		\cline{5-8}
		& & & &$\neq2$&& $\checkmark$&\textcolor{red}{Theorem 3.1,  \cite{HuangTGRSDCC2021} }\\
		\cline{3-8}
		& &$\neq 0$&$=0$& & &$\times$ &Theorem \ref{l=0not} (1)\\
		\cline{3-8}
		& &\multirow{2}{*}{$=0$}& & $=2$& &$\checkmark$&Theorem \ref{l=0yes} \\
		\cline{5-8}
		& & &&$\neq2$ & &$\times$&Theorem \ref{l=0not} (3)\\
		\cline{2-8}
		&\multirow{3}{*}{$2k+1$}&$\neq 0$&$\neq0$ & & & $\times$&\multirow{3}{*}{Theorem \ref{l=0not} (4)}\\
		\cline{3-7}
		& &$\neq 0$&$=0$ & & & $\times$& \\
		\cline{3-7}
		& &$=0$& & & & $\times$& \\
		\cline{2-8}
		&$\geq 2k+2$& & & & & $\checkmark$&\textcolor{red}{Theorem 6, \cite{ LiuTGRSDCC2021}}\\
		\hline
		\multirow{7}{*}{$1$}&$2k$& & & & & $\times$&Theorem \ref{notl=1n=2k+1yes2} (1)\\
		\cline{2-8}
		&\multirow{2}{*}{$2k+1$}& &$\neq 0$& & &$\checkmark$&Theorem \ref{l=1n=2k+1yes1}\\
		\cline{3-8}
		& & &$=0$& & &$\checkmark$&Theorem \ref{l=1n=2k+1yes2}\\
		\cline{2-8}
		&\multirow{2}{*}{$2k+2$}& & & & \makecell{$\left(0,\eta_2\right),$\\ $\eta_2\in\mathbb{F}_q^*$}& $\checkmark$& \textcolor{red}{Theorem 4.4 (2),  \cite{ZhangTGRSJAA2025}}\\
		\cline{3-8}
		& & & & &$\in\mathbb{F}_{q}^{2}\backslash\left\{\boldsymbol{0}\right\}$&$\checkmark$&Theorem \ref{lgeq1n=2k+l+1yes1}\\
		\cline{2-8}
		&$2k+3$& & & & & $\times$&Theorem \ref{notl=1n=2k+1yes2} (3)\\
		\cline{2-8}
		&$\geq 2k+4$& & & & & $\checkmark$&Theorem \ref{lgeq1ngeq2k+2l+2yes}\\
		\hline
		\multirow{4}{*}{$\geq 2$}&$2k$& & & & & $\times$&Theorem \ref{notl=1n=2k+1yes2} (1)\\ 
		\cline{2-8}
		&$2k+1$& & & & & $\times$&Theorem \ref{notl=1n=2k+1yes2} (2)\\ 
		\cline{2-8}
		&$2k+\ell+1$& & & & & $\checkmark$&Theorem \ref{lgeq1n=2k+l+1yes1}\\ 
		\cline{2-8}
		&$\geq 2k+2\ell+2$& & & & & $\checkmark$&Theorem \ref{lgeq1ngeq2k+2l+2yes}\\ 
		\hline
	\end{tabular}
\end{threeparttable} 
 
{\color{red}}
\end{document}